\documentclass[12pt]{imsart}

\usepackage[OT1]{fontenc}
\usepackage{url}
\usepackage{natbib,graphicx,amsmath,amssymb,fullpage,dsfont}
\usepackage{amsfonts}
\usepackage{fancybox}

\usepackage{lscape}
\usepackage{caption, subcaption}
\usepackage{geometry}

\usepackage[linesnumbered,ruled]{algorithm2e}

\usepackage{comment}
\usepackage{color}
\usepackage{imsart}
\usepackage[colorlinks,citecolor=blue,urlcolor=blue]{hyperref}

\usepackage{tikz,enumerate}
\usepackage{multirow,ctable}
\usepackage[normalem]{ulem}
\usepackage[framemethod=TikZ]{mdframed}
\usetikzlibrary{arrows}

\newdimen\arrowsize
\pgfarrowsdeclare{arcsq}{arcsq}
{
  \arrowsize=0.2pt
  \advance\arrowsize by .5\pgflinewidth
  \pgfarrowsleftextend{-4\arrowsize-.5\pgflinewidth}
  \pgfarrowsrightextend{.5\pgflinewidth}
}
{
  \arrowsize=1.5pt
  \advance\arrowsize by .5\pgflinewidth
  \pgfsetdash{}{0pt} 
  \pgfsetroundjoin   
  \pgfsetroundcap    
  \pgfpathmoveto{\pgfpoint{0\arrowsize}{0\arrowsize}}
  \pgfpatharc{-90}{-140}{4\arrowsize}
  \pgfusepathqstroke
  \pgfpathmoveto{\pgfpointorigin}
  \pgfpatharc{90}{140}{4\arrowsize}
  \pgfusepathqstroke
}

\usepackage{paralist}
\newtheorem{prop}{Proposition}

\newtheorem{example}{Example}

\newcommand{\D}{{\mathcal{D}}}

\def\minimize#1#2{  {\underset{#1}{\mathrm{minimize}}}\left\{#2\right\}}
\def\argmin#1#2{  {\underset{#1}{\mathrm{argmin}}}\left\{#2\right\}}

\newcommand\numberthis{\addtocounter{equation}{1}\tag{\theequation}}

\newcounter{xxx}
\usepackage{color}
\usepackage{ulem}

\begin{document}

\begin{frontmatter}
\title{Fast Nonconvex Deconvolution of Calcium Imaging Data}
\runtitle{Fast nonconvex deconvolution of calcium imaging Data}

\vspace{5mm}

\begin{aug}
\author{Sean Jewell\ead[label = e1]{swjewell@uw.edu}} 
\address{Department of Statistics \\ University of Washington \\ Seattle, Washington 98195 \\ \printead{e1}}

\author{Toby Dylan Hocking\ead[label = e3]{toby.hocking@mail.mcgill.ca}}
\address{Human Genetics Department \\ McGill University and Genome Qu\'ebec Innovation Centre\\ Montr\'eal, Qu\'ebec H3A 1A4 \\\printead{e3}}

\author{Paul Fearnhead\ead[label = e4]{p.fearnhead@lancs.ac.uk}}
\address{Department of Mathematics and Statistics \\ Lancaster University \\ Lancaster, UK LA1 4YF \\ \printead{e4}}

\author{Daniela Witten\ead[label = e2]{dwitten@uw.edu}}
\address{Departments of Statistics and Biostatistics \\ University of Washington \\ Seattle, Washington 98195 \\ \printead{e2}}

\runauthor{S. Jewell, D. Witten, T.D. Hocking, P. Fearnhead}

\end{aug}

\begin{abstract}

Calcium imaging data promises to transform the field of neuroscience by making it possible to record from large populations of neurons simultaneously. However, determining the exact moment in time at which a neuron spikes, from a calcium imaging data set, amounts to a non-trivial deconvolution problem which is of critical importance for downstream analyses. While a number of formulations have been proposed for this task in the recent literature, in this paper we focus on a formulation recently proposed in \cite{jewell2017exact} which has shown initial promising results. However, this proposal is slow to run on fluorescence traces of hundreds of thousands of timesteps. 

Here we develop a much faster online algorithm for solving the optimization problem of \cite{jewell2017exact} that can be used to deconvolve a fluorescence trace of 100,000 timesteps in less than a second. Furthermore, this algorithm overcomes a technical challenge of \cite{jewell2017exact} by avoiding the occurrence of so-called ``negative'' spikes. We demonstrate that this algorithm has superior performance relative to existing methods for spike deconvolution on calcium imaging datasets that were recently released as part of the \texttt{spikefinder} challenge (\url{http://spikefinder.codeneuro.org/}).


Our \texttt{C++} implementation, along with \texttt{R} and \texttt{python} wrappers, is publicly available on \texttt{Github} at \url{https://github.com/jewellsean/FastLZeroSpikeInference}.

\end{abstract}

\end{frontmatter}

\maketitle
 

\section{Introduction}
\label{sec:intro}

Due to recent advances in calcium imaging technology, it has become possible to record from large populations of neurons simultaneously in behaving animals \citep{dombeck2007imaging,ahrens2013whole,prevedel2014simultaneous}. These data result in a fluorescence trace for each neuron. 


However, most downstream analyses require not a fluorescence trace, but instead a measure of the neuron's activity over time. Consequently, a number of unsupervised and---more recently---supervised methods have been developed to infer neural activity on the basis of the fluorescence trace \citep{jewell2017exact, grewe2010high,pnevmatikakis2013bayesian,theis2016benchmarking,deneux2016accurate,sasaki2008fast,vogelstein2009spike,yaksi2006reconstruction,vogelstein2010fast,holekamp2008fast, friedrich2016fast, friedrichfast2017}. 


In this paper, we make use of a generative model that connects the observed fluorescence trace $y_{t}$ to the underlying and unobserved calcium concentration $c_{t}$, and the unknown spike times \citep{vogelstein2010fast, friedrich2016fast,friedrichfast2017}. This model assumes that the observed fluorescence is a noisy version of the underlying calcium, which exponentially decays, unless there is a spike, in which case there is an instantaneous increase in the calcium concentration. This leads directly to the model
\begin{align}
& y_t = \beta_0  + c_t + \epsilon_t, \quad \epsilon_t \sim_\text{ind.} (0, \sigma^2),  \quad t = 1, \ldots, T; \nonumber\\
& c_t = \gamma c_{t-1} + z_t, \quad t = 2,\ldots, T,
\label{eq:model}
\end{align}
where $z_t\geq 0$ denotes the potential spike in concentration. At most timesteps $z_t=0$, corresponding to no spike, and the calcium will decay exponentially at a rate governed by the parameter $\gamma$, which is assumed known. For simplicity, in what follows, we assume that the intercept $\beta_{0}$ is equal to zero. However, this is easy to relax; see Section~\ref{sec:generalizations}.

Under the additional assumption that the errors $\epsilon_{t}$ are normally distributed, model \eqref{eq:model} suggests estimating the concentration by solving the following constrained $\ell_{0}$ optimization problem
\begin{equation}
\minimize{c_1,\ldots,c_T, z_2,\ldots,z_T}{  \frac{1}{2} \sum_{t=1}^T \left( y_t -  c_t \right)^2 + \lambda \sum_{t=2}^T 1_{\left( z_t \neq 0 \right) }} \mbox{ subject to } z_t = c_t - \gamma c_{t-1} \geq 0,
   \label{eq:nonconvex-pos}
   \end{equation}
where $\lambda$ is a non-negative tuning parameter that controls the tradeoff between how closely the calcium concentration matches the fluorescence trace $(\sum_{t=1}^T(y_t - c_t)^2)$ and the number of non-zero spikes $(\sum_{t=2}^T1_{(z_t \neq 0)})$. The solution to this optimization problem directly provides an estimate for the spike times; that is, if $\hat{z}_t\neq 0$, then we infer a spike at time $t$. We note that this problem is over-parameterized, in the sense that knowing $c_1, \ldots, c_{T}$ determines $z_2, \ldots, z_{T}$. 

While \eqref{eq:nonconvex-pos} follows naturally from the biological process described in \eqref{eq:model}, the $\ell_{0}$ penalty makes the problem nonconvex and thus seemingly intractable. Consequently, rather than solving \eqref{eq:nonconvex-pos}, previous approaches have solved a convex relaxation to \eqref{eq:nonconvex-pos} \citep{vogelstein2010fast, friedrich2016fast,friedrichfast2017}, where the $\ell_0$ penalty is replaced by an $\ell_1$ penalty; see \eqref{eq:convex-fried}.

In recent work, \cite{jewell2017exact} showed that it is possible to efficiently solve the related nonconvex optimization problem 
 \begin{equation}
\minimize{c_1,\ldots,c_T,z_2,\ldots,z_T}{  \frac{1}{2} \sum_{t=1}^T \left( y_t - c_t \right)^2 + \lambda \sum_{t=2}^T 1_{\left( z_{t} \neq 0\right) }} \mbox{ subject to } z_t = c_t - \gamma c_{t-1},
   \label{eq:nonconvex-nopos}
   \end{equation}
which results from removing the positivity constraint, $c_{t} - \gamma c_{t-1} \geq 0$, from \eqref{eq:nonconvex-pos}. The positivity constraint enforces the biological property that a firing neuron can only cause the calcium concentration to increase (and never decrease).  Nonetheless, despite the slight loss in physical interpretability caused by the omission of the positivity constraint, \cite{jewell2017exact} showed that solving \eqref{eq:nonconvex-nopos} leads to improved performance over existing deconvolution approaches that perform a convex relaxation of \eqref{eq:nonconvex-pos}. In particular, while existing deconvolution approaches provide an adequate estimate of a neuron's firing rate over time, the method of \cite{jewell2017exact} provides an accurate estimate of the \textit{specific timesteps at which a neuron fires}. 

Unfortunately, the algorithm proposed in \cite{jewell2017exact} for solving \eqref{eq:nonconvex-nopos} is too slow to conveniently run on large-scale data. For traces of 100,000 timesteps, the \texttt{R} implementation runs on a laptop computer in a few minutes for a single value of the tuning parameter $\lambda$; in practice the user must apply the algorithm over a fine grid of values of $\lambda$, leading potentially to hours of computation time for a single trace. Furthermore, a single experiment could result in hundreds or thousands of fluorescence traces \citep{ahrens2013whole, vladimirov2014light}.

In this paper, we develop a fast algorithm for solving problem \eqref{eq:nonconvex-nopos}; for traces of 100,000 timesteps our \texttt{C++} implementation runs on a laptop computer in less than a second. Furthermore, this new algorithm can easily accommodate the positivity constraint that was omitted from \eqref{eq:nonconvex-nopos}; in other words, we can directly solve problem \eqref{eq:nonconvex-pos}. 

In what follows, we introduce our new algorithm for solving  \eqref{eq:nonconvex-pos} and  \eqref{eq:nonconvex-nopos} in Section~\ref{sec:algorithm}. We compare its performance in Section~\ref{sec:real-data} to a very popular approach for fluorescence trace deconvolution on a number of calcium imaging datasets that were recently released as part of the \texttt{spikefinder} challenge (\url{http://spikefinder.codeneuro.org/}). We close with a discussion in Section~\ref{sec:discussion}. The Appendix contains proofs, implementation considerations, pseudo-code for \eqref{eq:nonconvex-pos}, and an additional example.


\section{A fast functional pruning algorithm for solving problems \eqref{eq:nonconvex-nopos} and \eqref{eq:nonconvex-pos}}
\label{sec:algorithm}



\subsection{A review of \cite{jewell2017exact}}

\label{subsec:cp}

\cite{jewell2017exact} point out that the $\ell_0$ optimization problem \eqref{eq:nonconvex-nopos} is equivalent to a changepoint detection problem,
 \begin{equation}
	\minimize{0= \tau_0 < \tau_1<\ldots<\tau_k < \tau_{k+1} = T,k}{\sum_{j=0}^{k}  
 \D(y_{(\tau_{j}+1):\tau_{j+1}}) + \lambda k },
 \label{eq:cp}
 \end{equation}
where 
 \begin{equation}
 \label{eq:D}
 \D(y_{a:b}) \equiv \underset{\alpha}{\min}
  \left\{  \frac{1}{2} \sum_{t=a}^{b} \left( y_t - \alpha \gamma^{t-b}\right)^2 \right\}.
  \end{equation}
In problem \eqref{eq:cp}, we select the optimal changepoints $\tau_{1}, \ldots, \tau_{k}$ and the number of changepoints $k$ such that the overall cost of segmenting the data into $k+1$ exponentially decaying regions is as small as possible, where \eqref{eq:D} is the cost associated with the region that spans the $a$th to $b$th timesteps. Problems \eqref{eq:cp} and \eqref{eq:nonconvex-nopos} are equivalent in the sense that $\hat{z}_{\hat{\tau}_{1} + 1}\neq 0, \ldots, \hat{z}_{\hat{\tau}_{k} + 1} \neq 0$ and all other $\hat{z}_{t} = 0$.

To solve the changepoint problem, \cite{jewell2017exact} exploit a simple recursion \citep{jackson2005algorithm}, 
\begin{align*}
F(s) &= 
 \underset{0= \tau_{0}< \tau_{1}< \cdots < \tau_{k} < \tau_{k+1} = s, k}{\mathrm{min}} \left\{ \sum_{j=0}^{k}  
 \D(y_{(\tau_{j}+1):\tau_{j+1}}) + \lambda k \right\} = \underset{0\leq \tau < s}{\mathrm{min}}\left\{ F(\tau) + \D(y_{(\tau + 1):s}) + \lambda \right\}, \numberthis \label{eq:op-recursion}
\end{align*}
where $F(s)$ is the optimal cost of segmenting the data $y_{1:s}\equiv [y_{1}, \ldots, y_{s}]$, and where we define $F(0) \equiv -\lambda$. This results in an algorithm with computational complexity $\mathcal{O}(T^{2})$, which can be substantially improved by noticing that the minimization on the right hand side of \eqref{eq:op-recursion} can be performed over a smaller set $\mathcal{E}_{s}$ without sacrificing the global optimum \citep{killick2012optimal}; details are provided in \cite{jewell2017exact}. As mentioned in the introduction, the algorithm runs in a few minutes for traces of length 100,000, and yields the global optimum to \eqref{eq:nonconvex-nopos}. We note that the recursion \eqref{eq:op-recursion} does not naturally lead to an algorithm to solve \eqref{eq:nonconvex-pos}; this is discussed in further detail in Section~\ref{sec:fpop-minless}.


\subsection{Functional pruning for solving \eqref{eq:nonconvex-nopos}}
\label{sec:fpop}

\subsubsection{Motivation for functional pruning}
\label{sec:motivate-fpop}

\begin{figure}[htbp]
\begin{center}
\includegraphics[scale = 0.42]{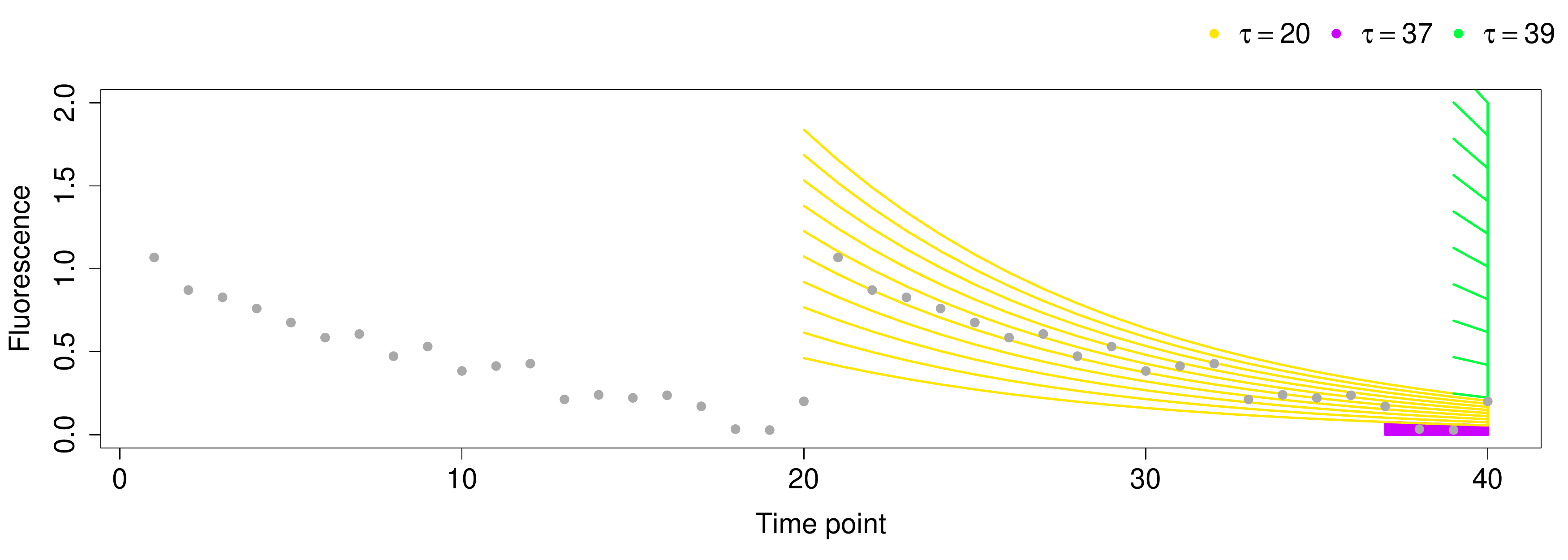}
\caption{A simple example to show that there are only a few possible values for the most recent changepoint before timestep 40.
We consider solving for the most recent changepoint given data $y_{1:40}$ and a range of possible values for the calcium concentration at the $40$th timestep, $c_{40}$. For a range of values of $c_{40}$, we display the estimated calcium concentration from the most recent changepoint before timestep 40. The colors indicate the time of the most recent changepoint. In this example there are only three possibilities for the most recent changepoint: $\{ 20, 37, 39\}$.}
\label{fig:motivate}
\end{center}
\end{figure}

In order to motivate the potential for a much faster algorithm for solving \eqref{eq:nonconvex-nopos} than the one proposed in \cite{jewell2017exact}, consider Figure~\ref{fig:motivate}.  

In this figure, we are interested in determining the optimal cost of segmenting the data up to time $40$, that is, calculating $F(40)$ in \eqref{eq:op-recursion}. Instead of directly applying the recursion \eqref{eq:op-recursion}, we consider a slightly different question: What is the best time for the most recent changepoint before the 40th timestep conditional on, and as a function of, the calcium concentration $c_{40}$? Conditional on the previously stored values $F(0), \ldots, F(39)$, and the data $y_{1:40}$, it is straightforward to calculate the best most recent changepoint for any value of the calcium concentration $c_{40}$.

Figure~\ref{fig:motivate} displays the most recent changepoint as a function of $c_{40}$. We observe that regardless of the value of the calcium at the current timestep --- and consequently, regardless of the fluorescence values $y_{41},y_{42},y_{43},\ldots,y_T$ --- \emph{the only possible times for the most recent changepoint before the 40th timestep are 20, 37, and 39}. 
  
However, the algorithm proposed in \cite{jewell2017exact} does not exploit the fact that 20, 37, and 39 are the only possible times for the most recent changepoint before the 40th timestep:  the minimization in \eqref{eq:op-recursion} is performed over the set $\{0,\ldots,39\}$, or else over a slightly smaller set $\{18, \ldots, 39\}$ using ideas from \cite{killick2012optimal}. This suggests that by viewing the cost of segmenting the data up until the $s$th timestep as a function of the \emph{calcium at the $s$th timestep}, we could potentially develop an algorithm that is much faster than the one in \cite{jewell2017exact} in that it would only require performing the minimization in \eqref{eq:op-recursion} over $\{20, 37, 39\}$. The idea of using this type of conditioning was first suggested by \cite{rigaill2015pruned} and \cite{maidstone2016optimal}, albeit to speed up algorithms for detecting changepoints in a different class of models.

\subsubsection{The functional pruning algorithm}
\label{sec:fpop-alg}

To begin, we substitute the cost function $\D(y_{(\tau + 1):s})$ into the recursion \eqref{eq:op-recursion}, in order to obtain
\begin{align}
F(s) &= \min_{0\leq \tau < s}\left\{F(\tau) + \D(y_{(\tau + 1):s}) + \lambda \right\} \nonumber \\
&= \min_{0\leq \tau < s}\left\{F(\tau) + \underset{\alpha}{\min}
 \left\{  \frac{1}{2} \sum_{t=\tau + 1}^{s} \left( y_t - \alpha \gamma^{t - s}\right)^2 \right\}
+ \lambda \right\}\nonumber \\
&= \underset{\alpha}{\min}  \min_{0\leq \tau < s}   \left\{F(\tau) +
 \left\{  \frac{1}{2} \sum_{t=\tau + 1}^{s} \left( y_t - \alpha \gamma^{t - s}\right)^2 \right\}
+ \lambda \right\} \nonumber \\
&=  \underset{\alpha}{\min}  \min_{0\leq \tau < s} \mathrm{Cost}_s^\tau(\alpha) \nonumber \\
&=  \underset{\alpha}{\min}  \mathrm{Cost}_s^*(\alpha), \label{eq:f-to-cost}
\end{align}
where
\begin{equation}
\mathrm{Cost}_s^\tau(\alpha) \equiv F(\tau) +
   \frac{1}{2} \sum_{t=\tau + 1}^{s} \left( y_t - \alpha \gamma^{t - s}\right)^2 + \lambda, \label{eq:cost-tau}
\end{equation}
and 
\begin{equation}
\mathrm{Cost}_s^*(\alpha) = \min_{0 \leq \tau < s} \mathrm{Cost}_s^\tau(\alpha). \label{eq:cost-star}
\end{equation}
In words, $\mathrm{Cost}_s^\tau(\alpha)$ is the cost of partitioning the data up until time $s$, given that the most recent changepoint was at time $\tau$, and the calcium at the $s$th timestep equals $\alpha$. $\mathrm{Cost}_s^*(\alpha)$ is the optimal cost of partitioning the data up until time $s$, given that the calcium at the $s$th timestep equals $\alpha$. 

The following proposition will prove useful in what follows.  
 \begin{prop}
 \label{prop:recursion}
 For $\mathrm{Cost}_{s}^*(\alpha)$ defined in \eqref{eq:cost-star}, the following recursion holds: 
\begin{equation}
\mathrm{Cost}_{s}^*(\alpha) = \min\left\{ \mathrm{Cost}_{s-1}^*(\alpha/\gamma), \min_{\alpha'} \mathrm{Cost}_{s-1}^*(\alpha') + \lambda \right\} + \frac12(y_s - \alpha)^2.
\label{eq:min-cost}
\end{equation}
\end{prop}
The proof of Proposition~\ref{prop:recursion} is in Appendix~\ref{sec:fun-proofs}. The recursion in \eqref{eq:min-cost} encompasses two possibilities: either there is a changepoint at the $(s-1)$st timestep, and we must determine the optimal cost up to that time $\left( \min_{\alpha'} \mathrm{Cost}_{s-1}^*(\alpha') + \lambda + \frac12(y_s - \alpha)^2\right)$, or there is no changepoint at the $(s-1)$st timestep $\left( \mathrm{Cost}_{s-1}^*(\alpha/\gamma) + \frac12(y_s - \alpha)^2\right)$. The recursion in \eqref{eq:min-cost} is reminiscent of \eqref{eq:op-recursion}, and raises the following question: can we use \eqref{eq:min-cost}
 as the basis for a recursive algorithm for solving the problem of interest, \eqref{eq:nonconvex-nopos}? At first, it appears almost hopeless, since the recursion \eqref{eq:min-cost} 
involves a \emph{function of $\alpha$}, a real-valued parameter. However, as we will see, it turns out that $\mathrm{Cost}_{s}^*(\alpha)$ and $\mathrm{Cost}_{s}^\tau(\alpha)$ are simple functions of  $\alpha$ that are easy to analytically manipulate.

Observe that, by definition \eqref{eq:cost-star}, the optimal cost $\mathrm{Cost}_{s}^{*}(\alpha)$ takes the form 
\begin{align} \label{eq:piecewise}
\mathrm{Cost}_{s}^{*}(\alpha) &= 
\begin{cases}
 \mathrm{Cost}_{s}^{0}(\alpha), & \alpha \in \mathcal{R}_{s}^{0} \\
 \vdots & \vdots \\
  \mathrm{Cost}_{s}^{s-1}(\alpha), & \alpha \in \mathcal{R}_{s}^{s-1}
\end{cases},
\end{align} 
where $\mathcal{R}_s^{\tau} \equiv \left\{ \alpha: \min_{0 \leq \tau' < s} \mathrm{Cost}_{s}^{\tau'}(\alpha) = \mathrm{Cost}_{s}^{\tau}(\alpha) \right\}$; this is the set of values for the calcium at the $s$th timestep such that the most recent changepoint occurred at time $\tau$.  Furthermore, by inspection of \eqref{eq:cost-tau}, we see that $\mathrm{Cost}_s^{\tau}(\alpha)$ is itself a quadratic function of $\alpha$ for all $\tau$.
Thus, $\mathrm{Cost}_{s}^{*}(\alpha)$  is in fact \emph{piecewise quadratic}. This means that in order to efficiently store the function $\mathrm{Cost}_{s}^{*}(\alpha)$, we must simply keep track of the regions $\mathcal{R}_s^{0},\ldots,\mathcal{R}_s^{s-1}$, as well as the three coefficients (constant, linear, quadratic) that define the quadratic function corresponding to each region. We will now present a small toy example illustrating how the recursion \eqref{eq:min-cost} can be used to build up optimal cost functions, each of which is piecewise quadratic. 

\begin{figure}[htbp]
\begin{center}
\begin{tabular}{lc}
 \begin{minipage}{0.07\textwidth}$s = 1:$\end{minipage}  &  \begin{minipage}{0.949\textwidth}\includegraphics[scale = 0.38]{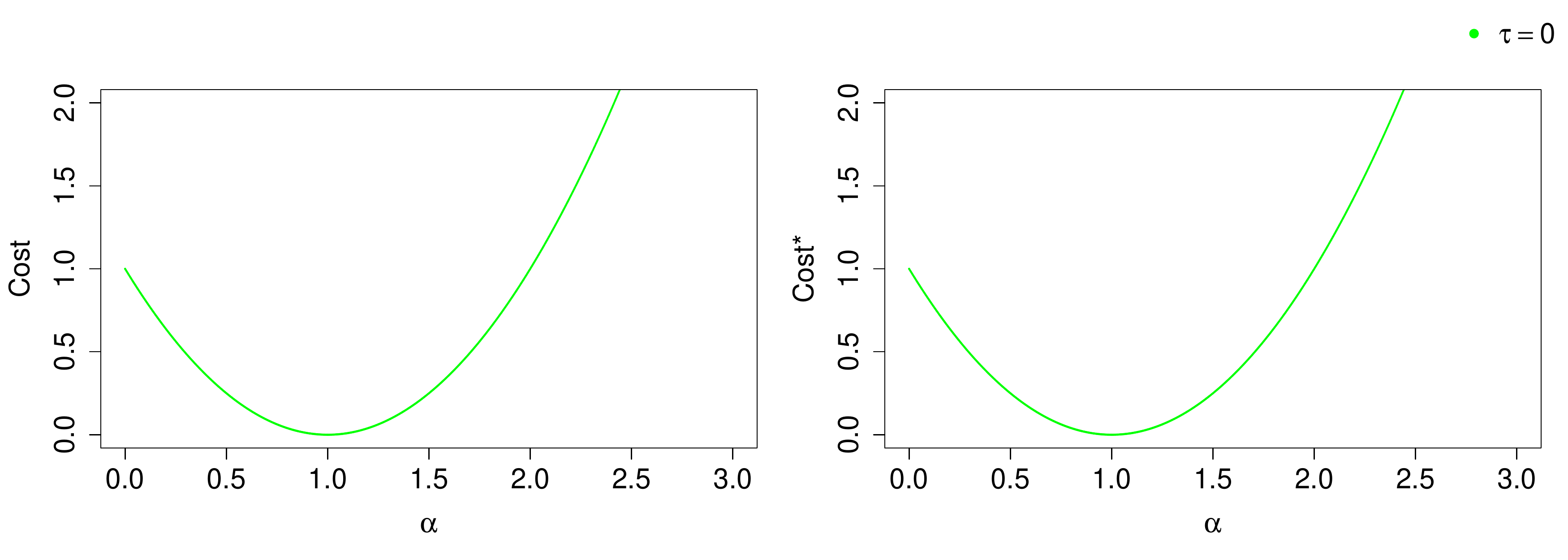}\end{minipage} \\
 \begin{minipage}{0.07\textwidth}$s = 2:$\end{minipage} & \begin{minipage}{0.949\textwidth}\includegraphics[scale = 0.38]{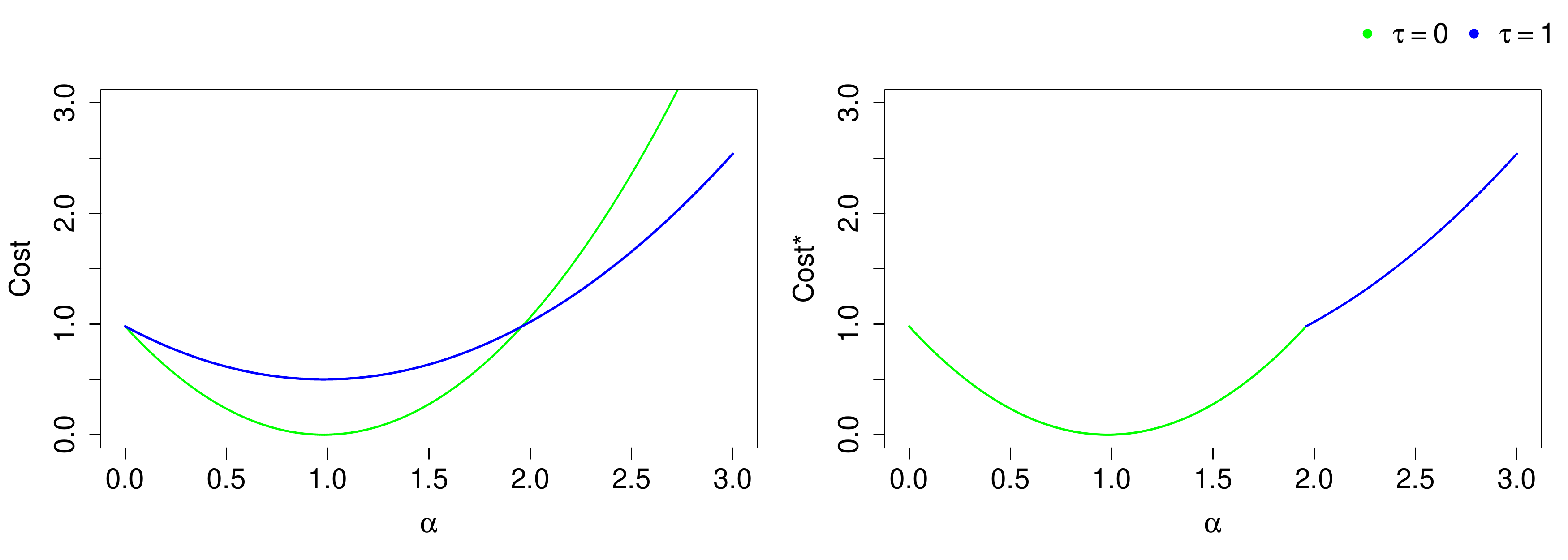} \end{minipage}\\
 \begin{minipage}{0.07\textwidth}$s = 3:$\end{minipage} & \begin{minipage}{0.949\textwidth}\includegraphics[scale = 0.38]{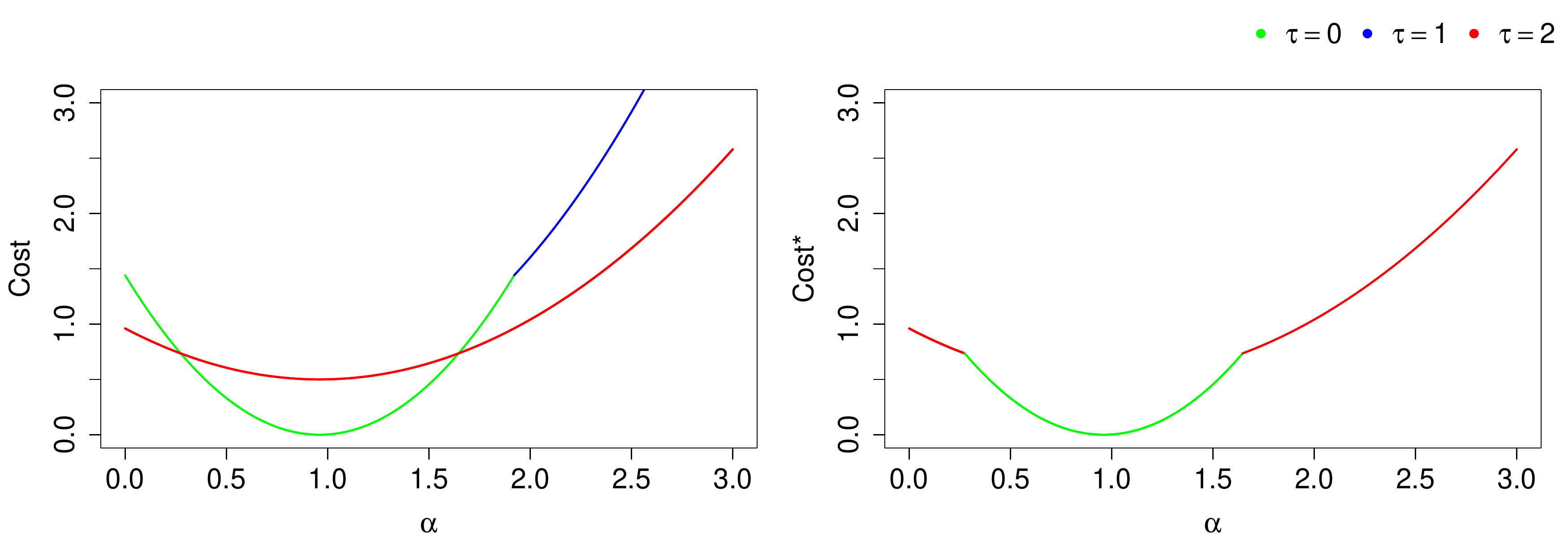}\end{minipage}\\
\end{tabular}
\caption{Evolution of $\mathrm{Cost}_{s}^{\tau}$ and $\mathrm{Cost}_{s}^{*}(\alpha)$ for Example~\ref{ex:recursion}. The left-hand panels display the functions $\mathrm{Cost}_{s-1}^*(\alpha/\gamma) + \frac12(y_s - \alpha)^2$ and $\min_{\alpha'} \mathrm{Cost}_{s-1}^*(\alpha') + \lambda  + \frac12(y_s - \alpha)^2$, and the right-hand panels show the function $\mathrm{Cost}_{s}^{*}(\alpha)$, which is the minimum of those two functions. Rows index the timesteps, $s = 1, 2, 3$. The functions are colored based on the timestep of the most recent changepoint, that is, the value of $\tau$ corresponding to $\mathcal{R}_s^{\tau}$. \emph{Top:} When $s=1$, $\mathrm{Cost}_1^*(\alpha)=\frac{1}{2}(1-\alpha)^2$; this corresponds to the region $\mathcal{R}_1^0 = [0, \infty)$. \emph{Center:} When $s=2$, $\mathrm{Cost}_2^*(\alpha)$ is the minimum of two quantities: $\mathrm{Cost}_{1}^*(\alpha/\gamma) + \frac12(y_2 - \alpha)^2$, which corresponds to the most recent changepoint being at timestep zero, and $\min_{\alpha'} \mathrm{Cost}_{1}^*(\alpha') + \lambda + \frac12(y_2 - \alpha)^2$, which corresponds to the most recent changepoint being at timestep one. These two functions are shown on the left-hand side, and $\mathrm{Cost}_2^*(\alpha$) is shown on the right-hand side. \emph{Bottom:} When $s=3$, $\mathrm{Cost}_3^*(\alpha)$ is calculated similarly; see Example~\ref{ex:recursion} for additional details.}
\label{fig:evolve}
\end{center}
\end{figure}

\begin{example} \label{ex:recursion}
Consider the simple dataset $y = [1.00, 0.98, 0.96, \ldots]$ with $\lambda = \frac12$ and $\gamma = 0.98$. We start with $\mathrm{Cost}_{1}^{*}(\alpha)$, which is just the quadratic centered around $y_{1}$, 
\begin{align*}
\mathrm{Cost}_{1}^{*}(\alpha) = \mathrm{Cost}_{1}^{0}(\alpha) = \frac12 (y_{1} - \alpha)^{2} = \frac12 (1.00 - \alpha)^{2}, \quad \alpha \in \mathcal{R}_{1}^{0} \equiv [0, \infty).
\end{align*}

Then, at the next time point, we form $\mathrm{Cost}_{2}^{*}(\alpha)$ based on \eqref{eq:min-cost},
\begin{align*}
\mathrm{Cost}_{2}^{*}(\alpha) &= 
\min\left\{ \mathrm{Cost}_{1}^*(\alpha/\gamma), \min_{\alpha'} \mathrm{Cost}_{1}^*(\alpha') + \lambda \right\} + \frac12(y_2 - \alpha)^2 \\
&= \min\left\{ \frac12(1 - \alpha / \gamma)^{2}, 0 + \frac12 \right\} + \frac12(0.98 - \alpha)^2 \\
&= 
\begin{cases}
\frac12( 1 - \alpha / \gamma)^{2} + \frac12(0.98 - \alpha)^2, &  \alpha \in \mathcal{R}_{2}^{0} \equiv [0, 2\gamma) \\
\frac12 + \frac12(0.98 - \alpha)^2, &  \alpha \in \mathcal{R}_{2}^{1} \equiv  [2\gamma, \infty)
\end{cases}.
\end{align*}

Again, using the recursion \eqref{eq:min-cost} we obtain the next optimal cost function,
{\footnotesize
\begin{align*}
\mathrm{Cost}_{3}^{*}(\alpha) &= 
\min\left\{ \mathrm{Cost}_{2}^*(\alpha/\gamma), \min_{\alpha'} \mathrm{Cost}_{2}^*(\alpha') + \lambda \right\} + \frac12(y_3 - \alpha)^2 \\
&= \min\left\{ \mathrm{Cost}_{2}^*(\alpha/\gamma), \frac12 \right\} + \frac12(0.96 - \alpha)^2 \\
&= 
\begin{cases}
\frac12 + \frac12(0.96 - \alpha)^2, &  \alpha \in \mathcal{R}_{3}^{2} \equiv \gamma^{2}\left\{\left[0,  1 - \frac{1}{\sqrt{1 + \gamma ^{ 2}}} \right) \cup \left[ 1 + \frac{1}{\sqrt{1 + \gamma ^{ 2}}}, \infty\right)\right\}\\
\frac12( 1 - \alpha / \gamma^{2})^{2} + \frac12(0.98 - \alpha / \gamma)^2 + \frac12(0.96 - \alpha)^2,&  \alpha \in \mathcal{R}_{3}^{0} \equiv   \gamma^{2}\left[ 1 - \frac{1}{\sqrt{1 + \gamma ^{ 2}}} ,  1 + \frac{1}{\sqrt{1 + \gamma ^{ 2}}} \right)
\end{cases}.
\end{align*}
}

We note that the $\mathrm{Cost}_{3}^{*}(\alpha)$ is defined over just $\mathcal{R}_{3}^{0}$ and $\mathcal{R}_{3}^{2}$. This example is displayed in Figure~\ref{fig:evolve}.

 \end{example}

Although we have shown how to efficiently build optimal cost functions $\mathrm{Cost}_{s}^{*}(\alpha)$ from $s = 1, \ldots, T$, it remains to establish that these cost functions can be used to determine the optimal changepoints, that is, the values of $\tau_{1}, \ldots, \tau_{k}$ that solve \eqref{eq:nonconvex-nopos}. Conceptually, we must find the value of $\tau$ that satisfies
\begin{align}
\tau^{*}(s) &= \{ \tau : \min_{\alpha}	\mathrm{Cost}_{s}^{\tau}(\alpha) = \min_{\alpha} \mathrm{Cost}_{s}^{*}(\alpha)		\}
\label{eq:decode}
\end{align} 
for $\tau^{*}(T), \tau^{*}(\tau^{*}(T)), \ldots$ until $0$ is obtained. Full details are provided in Algorithm~\ref{alg:fpop}. To summarize, we have developed a recursive algorithm for solving \eqref{eq:nonconvex-nopos} using the recursions in Proposition~\ref{prop:recursion}.

 \begin{example} \label{ex:recursion-revisit} 
 
``Example 1 revisited''

We return to Example~\ref{ex:recursion} to illustrate how \eqref{eq:decode} can be used to determine the optimal changepoints. In the interest of simplicity, we assume that $T = 3$; in other words, we have observed all of the data. Then, 
\begin{align*}
\tau^{*}(3) = \{ \tau : \min_{\alpha}	\mathrm{Cost}_{3}^{\tau}(\alpha) = \min_{\alpha} \mathrm{Cost}_{3}^{*}(\alpha)		\},
\end{align*}
where 

\begin{align*}
\min_{\alpha} \mathrm{Cost}_{3}^{\tau}(\alpha) &= 
\begin{cases}
\min_{\alpha}\mathrm{Cost}_{3}^{2}(\alpha) = 0.73, &  \alpha \in \mathcal{R}_{3}^{2} \\
\min_{\alpha}\mathrm{Cost}_{3}^{0}(\alpha) = 5.4\times 10^{-8}, &  \alpha \in \mathcal{R}_{3}^{0} 
\end{cases}.
\end{align*}

Therefore, the most recent changepoint is $\tau^{*}(3) = 0$. In fact, since the most recent changepoint is at timestep 0, we say that there are no changepoints. 
 
 \end{example}

Algorithm~\ref{alg:fpop} is an instance of the class of functional pruning algorithms proposed in \cite{maidstone2016optimal}. 
 
 \subsubsection{Computational time of functional pruning}
 
 We saw in Example~\ref{ex:recursion} that  Proposition~\ref{prop:recursion} can lead to a  recursive algorithm for solving the problem of interest \eqref{eq:nonconvex-nopos}.
  At first glance, since $\mathrm{Cost}_{s}^{*}(\alpha)$ is piecewise quadratic with $s$ regions \eqref{eq:piecewise}, and our recursive algorithm requires computing $\mathrm{Cost}_{1}^{*}(\alpha), \ldots, \mathrm{Cost}_{T}^{*}(\alpha)$, it appears that a total of $1+2+\ldots+T=O(T^2)$ operations must be performed in order to deconvolve a fluorescence trace of length $T$.
  Critically, however, this is not the case. 
  This is because, in practice, $\mathrm{Cost}_{s}^{*}(\alpha)$  is piecewise quadratic with \emph{substantially fewer than $s$ regions}, as we saw in Figure 1. To see this, 
  recall from \eqref{eq:piecewise} that the $\tau$th region up to timestep $s$ is defined as $\mathcal{R}_s^{\tau} \equiv \left\{ \alpha: \min_{0 \leq \tau' < s} \mathrm{Cost}_{s}^{\tau'}(\alpha) = \mathrm{Cost}_{s}^{\tau}(\alpha) \right\}$. However, if $\mathcal{R}_s^{\tau}$ is the empty set --- that is, if there is no $ \alpha$ such that $\min_{0 \leq \tau' < s} \mathrm{Cost}_{s}^{\tau'}(\alpha) = \mathrm{Cost}_{s}^{\tau}(\alpha) $ --- then $\mathrm{Cost}_{s}^{*}(\alpha)$  is, in fact, not a function of the $\tau$th region. 

   In practice, $\mathcal{R}_s^{\tau}$ will often be the empty set. For instance, see Figure~\ref{fig:motivate}. We note that in this example, at timestep $s=40$, the optimal cost function is only a function of three regions,
\begin{align*}
\mathrm{Cost}_{40}^*(\alpha) &= 
\begin{cases}
 1.88 \alpha ^{2} - 0.17 \alpha + 2.08, & \alpha \in \mathcal{R}_{40}^{37} \equiv [0, 0.06]  \\
 142.08 \alpha ^{2} - 39.60 \alpha + 3.85, & \alpha \in \mathcal{R}_{40}^{20} \equiv [0.06, 0.22] \\
 0.50 \alpha^{2} - 0.10 \alpha + 2.10, & \alpha \in \mathcal{R}_{40}^{39} \equiv [0.22, \infty) \\
\end{cases}.
\end{align*}
In a similar way, in Example~\ref{ex:recursion}, we saw that $\mathrm{Cost}_{3}^{*}(\alpha)$ was a function of two regions. 

\begin{figure}[h!]
\begin{center}
\includegraphics[width = \textwidth]{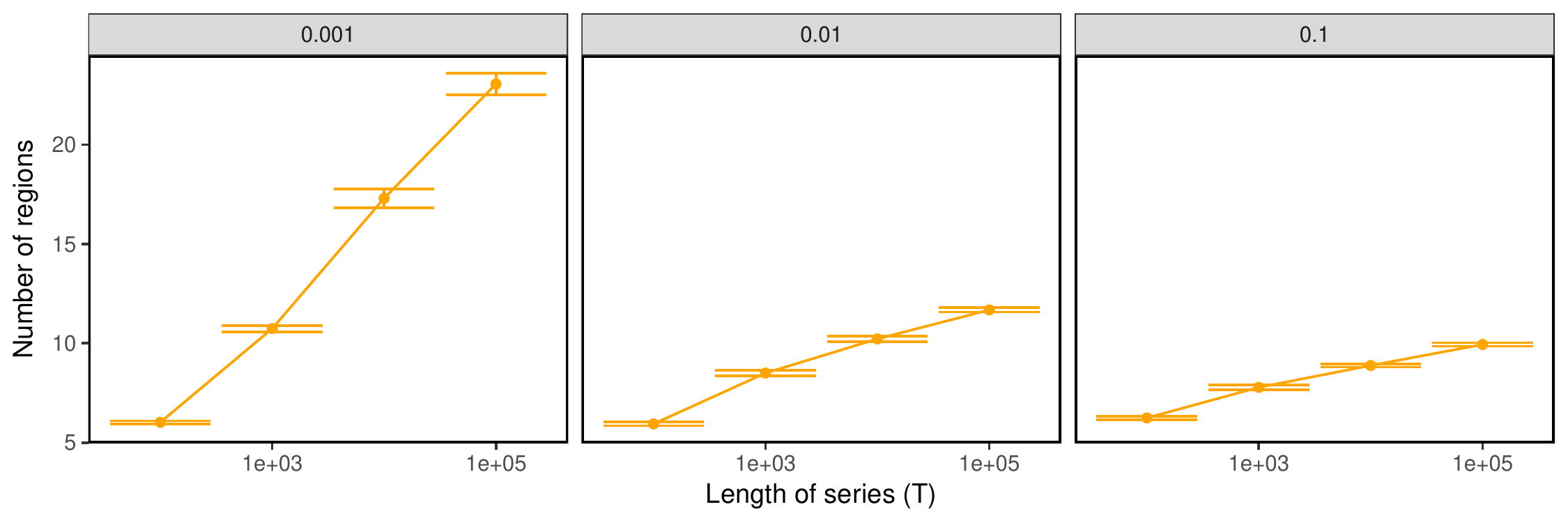}
\caption{Maximum number of regions from solving \eqref{eq:nonconvex-nopos} with $\lambda = 1$. Fifty sample datasets are simulated according to \eqref{eq:model} with coefficient $\beta_{0} = 0$, decay parameter $\gamma = 0.998$, normal errors $\epsilon_{t} \overset{\mathrm{ind}}{\sim} \mathrm{N}(0, \sigma = 0.15)$, and Poisson distributed spikes $s_{t} \overset{\mathrm{ind}}{\sim} \mathrm{Pois}(\theta)$ where $\theta \in \{ 0.1, 0.01, 0.001\}$. Panels correspond to different values of $\theta$.}
 \label{fig:regions}
\end{center}
\end{figure}

Therefore, though its worst-case performance is upper-bounded by $O(T^2)$, in practice, Algorithm~\ref{alg:fpop} is typically \emph{much} faster than this. Figure~\ref{fig:regions} illustrates that the maximum number of regions is a small fraction of the series length; for series of length $100,000$, fewer than 30 regions are required. 

\begin{figure}[h!]
\begin{center}
\includegraphics[scale = 0.4]{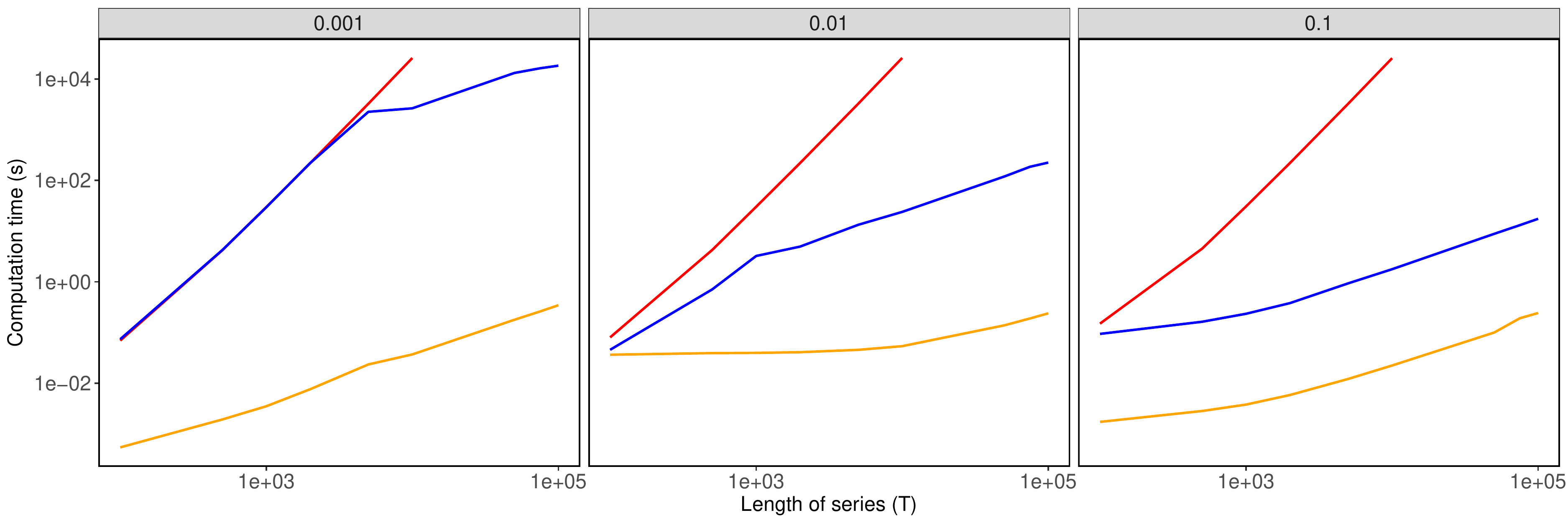}
\caption{Timing comparisons between three algorithms for solving \eqref{eq:nonconvex-nopos} with $\lambda = 1$. Functional pruning approach used in Algorithm~\ref{alg:fpop} (orange), and two algorithms from \cite{jewell2017exact}: one based on recursion \eqref{eq:op-recursion} (red), and one based on an improvement to \eqref{eq:op-recursion} called inequality pruning that makes use of ideas from \cite{killick2012optimal} (blue). Fifty sample datasets are simulated according to \eqref{eq:model} with coefficient $\beta_{0} = 0$, decay parameter $\gamma = 0.998$, normal errors $\epsilon_{t} \overset{\mathrm{ind}}{\sim} \mathrm{N}(0, \sigma = 0.15)$, and Poisson distributed spikes $s_{t} \overset{\mathrm{ind}}{\sim} \mathrm{Pois}(\theta)$ where $\theta \in \{ 0.1, 0.01, 0.001\}$. Standard errors are on average $< 0.1\%$ of the average computation time. Panels correspond to different values of $\theta$. Timing results were obtained on a Intel Xeon E5-2620 2.0 GHz processor.}
 \label{fig:timing}
\end{center}
\end{figure}

Furthermore, by slightly modifying Theorem 6.1 of \cite{maidstone2016optimal}, we can show that Algorithm~\ref{alg:fpop} is no worse than the algorithm proposed in \cite{jewell2017exact}. In fact, as shown in Figure~\ref{fig:timing}, Algorithm~\ref{alg:fpop} is typically up to a thousand times faster than that of \cite{jewell2017exact} on a fluorescence trace of length 100,000. In simulations, our \texttt{C++} implementation of Algorithm~\ref{alg:fpop} runs in less than one second on traces of length 100,000. 

\begin{algorithm}[h!]
    \SetKwInOut{Initialize}{Initialize}
    \SetKwInOut{Output}{Output}
	\Initialize{Compute $\mathrm{Cost}_{1}^{*}(\alpha) := \frac12 (y_{1} - \alpha)^{2}$}
	\ForEach{timestep $s = 2, \ldots, T$}{
		Calculate and store $\mathrm{Cost}_{s}^{*}(\alpha) := \min\{ \mathrm{Cost}_{s-1}^{*}(\alpha / \gamma), \min_{\alpha'}\mathrm{Cost}_{s-1}^{*}(\alpha') + \lambda\} + \frac12 (y_{s} - \alpha)^{2}$}
		
		// Use optimal cost functions to determine changepoints $\tau_{1}, \ldots, \tau_{k}$\\
		Initialize list of changepoints $cp := (T)$ \\
		Set the current changepoint $\tau^{cur} := T$ \\
		Initialize list of estimated calcium concentrations $c := ()$ \\
			\While{$\tau^{cur} > 0$}{
$\tau^{prev} := \tau^{cur}$ \\
Determine the most recent changepoint $\tau^{cur} := \{ \tau : \min_{\alpha}	\mathrm{Cost}_{\tau^{prev}}^{\tau}(\alpha) = \min_{\alpha} \mathrm{Cost}_{\tau^{prev}}^{*}(\alpha)		\}$ \\
Determine the calcium concentration at $\tau^{prev}$, $\alpha^{*} := \argmin{\alpha}{\mathrm{Cost}_{\tau^{prev}}^{\tau^{cur}}(\alpha)} $ \\
Update list of calcium concentrations, $ c := (\alpha^{*}, c)$ \\
\ForEach{timestep $s = (\tau^{prev} - 1), \ldots, (\tau^{cur} + 1)$}{
Calculate calcium concentration, $\alpha^{*} / \gamma$, and then append to list, $c := (\alpha^{*} / \gamma, c)$ \\
Scale $\alpha^{*} := \alpha^{*} / \gamma$ \\
}

Update list of changepoints $cp := (\tau^{cur}, cp)$ \\
		} 
    \Output{Set of changepoints $cp$, number of changepoints $k := \mathrm{card}(cp)$, and estimated calcium concentrations $c$.
    } 
\caption{A functional pruning algorithm for solving \eqref{eq:nonconvex-nopos} }\label{alg:fpop}
\end{algorithm}

\subsection{An efficient algorithm to solve the constrained problem \eqref{eq:nonconvex-pos}} 
\label{sec:fpop-minless}

As stated in the introduction, our main interest is to solve \eqref{eq:nonconvex-pos} for the global optimum. Problem \eqref{eq:nonconvex-pos} differs from problem \eqref{eq:nonconvex-nopos} in that there is an additional constraint that enforces biological reality: firing neurons can only cause an increase, but not a decrease, in the calcium concentration. The algorithm in \cite{jewell2017exact} cannot be used to solve \eqref{eq:nonconvex-pos}, because it relies on the recursion in \eqref{eq:op-recursion}, which does not allow for any dependence in the calcium concentration before and after a changepoint. Thus, at the time of this writing, there are no algorithms available to efficiently solve \eqref{eq:nonconvex-pos} for the global optimum. 


In this section we utilize a simple modification, due to \cite{hocking2017log}, to the functional recursion \eqref{eq:min-cost} that ensures that the constraint $c_{t} - \gamma c_{t-1} \geq 0$ is satisfied. First, recall from \eqref{eq:min-cost} that 
\begin{align*}
\mathrm{Cost}_{s}^{*}(\alpha)&= \min\left\{ \mathrm{Cost}_{s-1}^*(\alpha/\gamma), \min_{\alpha'} \mathrm{Cost}_{s-1}^*(\alpha') + \lambda \right\} + \frac12(y_s - \alpha)^2,
\end{align*}
where we take the minimum over two terms, which result from adding an additional point $y_{s}$ to the current segment $\left(\mathrm{Cost}_{s-1}^*(\alpha/\gamma) + \frac12(y_s - \alpha)^2 \right)$, and adding a new candidate changepoint at $s-1$ and starting a new segment at timestep $s$ $\left( \min_{\alpha'} \mathrm{Cost}_{s-1}^*(\alpha') + \lambda  + \frac12(y_s - \alpha)^2 \right)$.

In the latter case, if there is a spike at the $s$th timestep, then in order to enforce the positivity constraint, $z_s = c_s - \gamma c_{s-1} \geq 0$, the term $\min_{\alpha'} \mathrm{Cost}_{s-1}^*(\alpha') + \lambda$ in \eqref{eq:min-cost} needs to be modified to 
\begin{align}
\min_{\alpha \geq \alpha'} \mathrm{Cost}_{s-1}^*(\alpha' / \gamma) + \lambda.
\label{eq:intro-min-less}
\end{align}

Therefore, we replace \eqref{eq:min-cost} with 
\begin{align}
\mathrm{Cost}_{s}^{*}(\alpha)&= \min\left\{ \mathrm{Cost}_{s-1}^*(\alpha/\gamma), \min_{\alpha\geq \alpha'} \mathrm{Cost}_{s-1}^*(\alpha' / \gamma) + \lambda \right\} + \frac12(y_s - \alpha)^2,
\label{eq:min-less-cost}
\end{align}
and, in a slight abuse of notation, define the cost function associated with problem \eqref{eq:nonconvex-pos} as
\begin{align}
\mathrm{Cost}_s^\tau(\alpha) \equiv \min_{\alpha \geq \alpha'} \left[ \mathrm{Cost}_{\tau}^{*}(\alpha' / \gamma) +
   \frac{1}{2} \sum_{t=\tau + 1}^{s} \left( y_t - \alpha \gamma^{t - s}\right)^2 + \lambda\right]. \label{eq:cost-tau-constrain}
\end{align}
Equations \eqref{eq:min-less-cost} and \eqref{eq:cost-tau-constrain} can be used to develop an efficient recursive algorithm to solve problem \eqref{eq:nonconvex-pos}. Details of the algorithm itself are included in Appendix~\ref{sec:fast-algo-min-less}. A continuation of Example~\ref{ex:recursion} that solves \eqref{eq:nonconvex-pos} is included in Appendix~\ref{sec:min-less-example}.

\subsection{Solving \eqref{eq:model} for non-zero intercept $\beta_{0}$} 
\label{sec:generalizations} 

Thus far, we have considered \eqref{eq:model} with $\beta_{0} = 0$. To accommodate the possibility of nonzero baseline calcium, we consider the problem
  \begin{equation}
\minimize{c_1,\ldots,c_T,s_2,\ldots,s_T, \beta_{0}}{  \frac{1}{2} \sum_{t=1}^T \left( y_t - (\beta_{0}+ c_t) \right)^2 + \lambda \sum_{t=2}^T 1_{\left( s_{t} \neq 0\right) }} \mbox{ subject to } s_t \geq c_t - \gamma c_{t-1}.
   \label{eq:linear-extension}
   \end{equation}
Instead of directly solving \eqref{eq:linear-extension} with respect to $(c_{1}, \ldots, c_{T}, s_{2}, \ldots, s_{T}, \beta_{0})$, we consider a fine grid of values for $\beta_0$, and solve \eqref{eq:nonconvex-pos} with $y-\beta_0$ using Algorithm~\ref{alg:fpop-minless}, for each value of $\beta_0$ considered. The solution to \eqref{eq:linear-extension} is the set $\{\hat{c}_1,\ldots,\hat{c}_T,\hat{s}_2,\ldots,\hat{s}_T,\beta_0\}$ corresponding to the value of $\beta_0$ that led to the the smallest value of the objective, over all values of $\beta_0$ considered.


\section{Real data experiments}
\label{sec:real-data}

In this section, we illustrate the performance of the solution to \eqref{eq:nonconvex-pos} for spike deconvolution across a number of datasets, which were aggregated and standardized as part of the recent \texttt{spikefinder} challenge (\url{http://spikefinder.codeneuro.org/}). Each dataset consists of both calcium and electrophysiological recordings for a single cell; we will treat the spikes ascertained using electrophysiological recording as the ``ground truth'', and will quantify the ability of spike deconvolution algorithms to recover these ground truth spikes on the basis of the calcium recordings. The data sets differ in terms of the choice of calcium indicator (GCaMP5, GCaMP6, jRCAMP, jRGECO, OGB), scanning technology (AOD, galvo, and resonant), and circuit under investigation (V1 and retina).


Throughout this section, we compare our proposal \eqref{eq:nonconvex-pos} to a recent proposal from the literature that employs an $\ell_{1}$ (convex) relaxation to \eqref{eq:nonconvex-pos},
\begin{equation}
\minimize{c_1,\ldots,c_T, z_2,\ldots,z_T}{  \frac{1}{2} \sum_{t=1}^T \left( y_t - c_t \right)^2 + \lambda |c_{1}| + \lambda \sum_{t=2}^T | z_t |} \mbox{ subject to } z_t = c_t - \gamma c_{t-1} \geq 0, 
   \label{eq:convex-fried}
   \end{equation}
 proposed by \cite{friedrich2016fast} and \cite{friedrichfast2017}. We use the OASIS package to solve \eqref{eq:convex-fried} (\url{https://github.com/j-friedrich/OASIS}). Since the solution to \eqref{eq:convex-fried} often results in many small non-zero elements of $\hat{z}_{t}$, we consider post-thresholding. That is, given $\hat{z}_{2}, \ldots, \hat{z}_{T}$ that solve \eqref{eq:convex-fried}, and a threshold $L>0$, we set $\tilde{z}_{t} = \hat{z}_{t} 1_{(\hat{z}_{t} > L)}$; in other words, we only conclude that a spike is present if $\hat{z}_{t} > L$. 

In Section~\ref{sec:comparisons} we compare our proposed approach, \eqref{eq:nonconvex-pos}, and the proposal of \cite{friedrich2016fast} and \cite{friedrichfast2017}, \eqref{eq:convex-fried}, on data from the \texttt{spikefinder} challenge. We describe our experimental approach in Section~\ref{sec:exp-methods-desc}. Section~\ref{sec:single-cell} illustrates these methods for a single cell, and in Section~\ref{sec:results-spikefinder}, we examine results for all datasets considered in the \texttt{spikefinder} challenge. In Section~\ref{sec:no-neg-spikes}, we illustrate on a real-data example that solving \eqref{eq:nonconvex-pos} gives superior estimates than solving \eqref{eq:nonconvex-nopos}. In Section~\ref{sec:calibrate}, we make a connection between the instantaneous increase in calcium concentration to a spike in \eqref{eq:nonconvex-pos} and the actual number of recorded spikes.

\subsection{Comparison of \eqref{eq:nonconvex-pos} to \eqref{eq:convex-fried} on data from the \texttt{spikefinder} challenge}
\label{sec:comparisons}
\subsubsection{Description of methods for Sections~\ref{sec:single-cell}--\ref{sec:results-spikefinder}}
\label{sec:exp-methods-desc}

We now describe the methods that will be used in Sections~\ref{sec:single-cell}--\ref{sec:results-spikefinder}. Our main objective is to accurately estimate the number of spikes and the times at which spikes occur. Thus, we use two measures that directly compare two spike trains, both of which have been used extensively in the neuroscience literature \citep{quiroga2009extracting, reinagel2000temporal, gerstner2014neuronal} (i)  van Rossum distance \citep{van2001novel, houghton2012efficient}; and (ii) Victor-Purpura distance \citep{victor1997metric,victor1996nature}. We also use an additional measure: (iii) the correlation between two downsampled spike trains; details of this measure are provided in \cite{theis2016benchmarking}. As we will see, measures (i) and (ii) are sensitive to the number and timing of spikes, whereas measure (iii) is somewhat insensitive to the number and timing of the spikes, and instead quantifies the similarity between the spike rates. 

To analyze the performances of the proposals \eqref{eq:nonconvex-pos} and \eqref{eq:convex-fried} over a single fluorescence trace, we take a training/test set approach. Given a fluorescence trace of length $T$, the first $\lfloor T/2\rfloor $ timesteps are used in the training set, and the remainder are used for the test set. We solve \eqref{eq:nonconvex-pos} and \eqref{eq:convex-fried} for a range of values of the tuning parameter $\lambda$ on the training set; in the case of \eqref{eq:convex-fried} we also use a range of threshold values $L$. 

As pointed out by \cite{pachitariu2017robustness}, estimating the decay rate $\gamma$ in \eqref{eq:model} is difficult. Therefore, as in \cite{pachitariu2017robustness}, we categorize calcium indicators into three groups based on their decay properties. As in  \cite{vogelstein2010fast}, within each calcium indicator rate category, we set $\gamma = 1 - \frac{\Delta}{\phi}$, where $\Delta$ is 1 / (frame rate), and $\phi$ is a time-scale parameter based on the category, defined as 
\begin{align*}
\phi &= 
\begin{cases}
 0.7, & \mbox{ fast category} \\
  1.25, & \mbox{ medium category} \\
   2, & \mbox{ slow category} 
\end{cases}.
\end{align*}
For example, in Figure~\ref{fig:example-distances-comparison}, GCaMP6f is classified as a fast indicator and the data is recorded at 100Hz. Therefore, we take $\gamma \approx 0.986$. 

For all tuning parameter values considered, we apply the three measures mentioned earlier to the estimated and true spike trains, and select the tuning parameter values that optimize these measures. We then apply \eqref{eq:nonconvex-pos} and \eqref{eq:convex-fried} to the test set with the selected values of the tuning parameters, and evaluate test set performance using the three measures described earlier. 

\subsubsection{Results for a single cell}
\label{sec:single-cell}

In Figure~\ref{fig:example-distances-comparison}, we illustrate this procedure for cell 13, GCaMP6f, V1, from \cite{chen2013ultrasensitive}. Each row corresponds to one of the measures described in Section~\ref{sec:exp-methods-desc}. The left column displays these measures on the training set, for the solution to \eqref{eq:nonconvex-pos} with different values of $\lambda$, and for the post-thresholded solution to \eqref{eq:convex-fried} with different values of $\lambda$ and $L$. The right column shows the fluorescence trace along with the estimated spikes, on the test set, using tuning parameters selected on the training set. 

There are a number of important observations to draw from Figure~\ref{fig:example-distances-comparison}. As measured by van Rossum and Victor-Purpura, the estimated spikes from \eqref{eq:nonconvex-pos} are much more accurate than those estimated (and post-thresholded) using the convex relaxation \eqref{eq:convex-fried}. This agrees with our visual inspection of the right hand panel: the estimated spikes from problem \eqref{eq:nonconvex-pos} more closely match the number and timings of the true spikes than those estimated from problem \eqref{eq:convex-fried}.

By contrast, if performance is measured by correlation, then the estimated spikes obtained from \eqref{eq:convex-fried} result in slightly better performance than the estimated spikes from \eqref{eq:nonconvex-pos}. However, in the training set there are 75 true spikes, whereas \eqref{eq:convex-fried} outperforms \eqref{eq:nonconvex-pos} when approximately 200 spikes are estimated. Therefore, selecting the tuning parameter for \eqref{eq:convex-fried} based on correlation leads to a \emph{substantial overestimate} of the number of spikes, and therefore poor overall accuracy in the number and timing of the spikes. This pattern has been observed in other $\ell_{1}$ regularization problems  \citep{zou2006adaptive, maidstone2017detecting}. 

To summarize, when van Rossum and Victor-Purpura distance are used to evaluate performance, our proposal \eqref{eq:nonconvex-pos} substantially outperforms the approach in \eqref{eq:convex-fried}. When performance is evaluated using correlation, the performance of \eqref{eq:convex-fried} is slightly better than that of \eqref{eq:nonconvex-pos}; however, this better performance is achieved when far too many spikes are estimated, indicating that correlation is a poor choice for quantifying the accuracy of spike detection.

\begin{landscape}
\newgeometry{left=0.1cm,bottom=0.1cm} 
\begin{figure}[h!]
\begin{center}
\includegraphics[scale = 0.35]{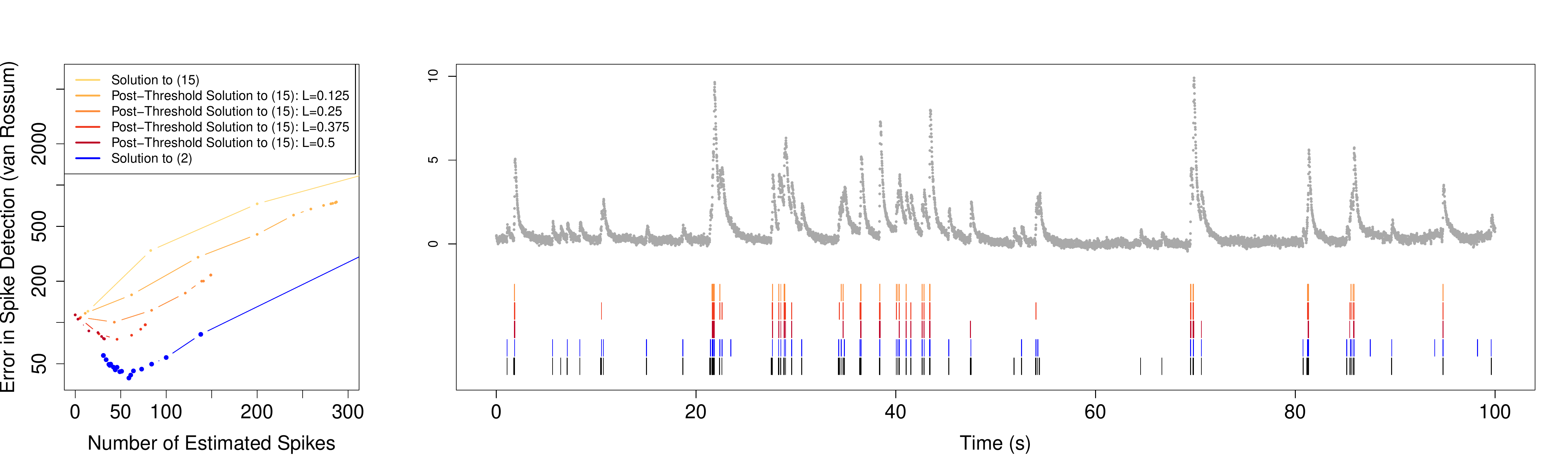}
\includegraphics[scale = 0.35]{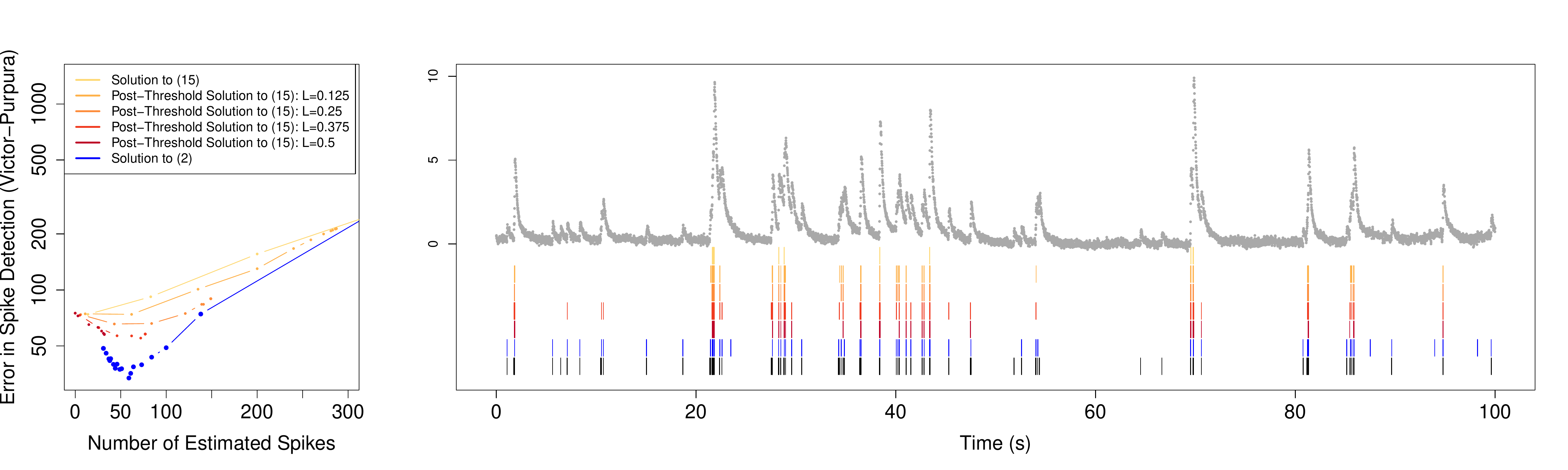}
\includegraphics[scale = 0.35]{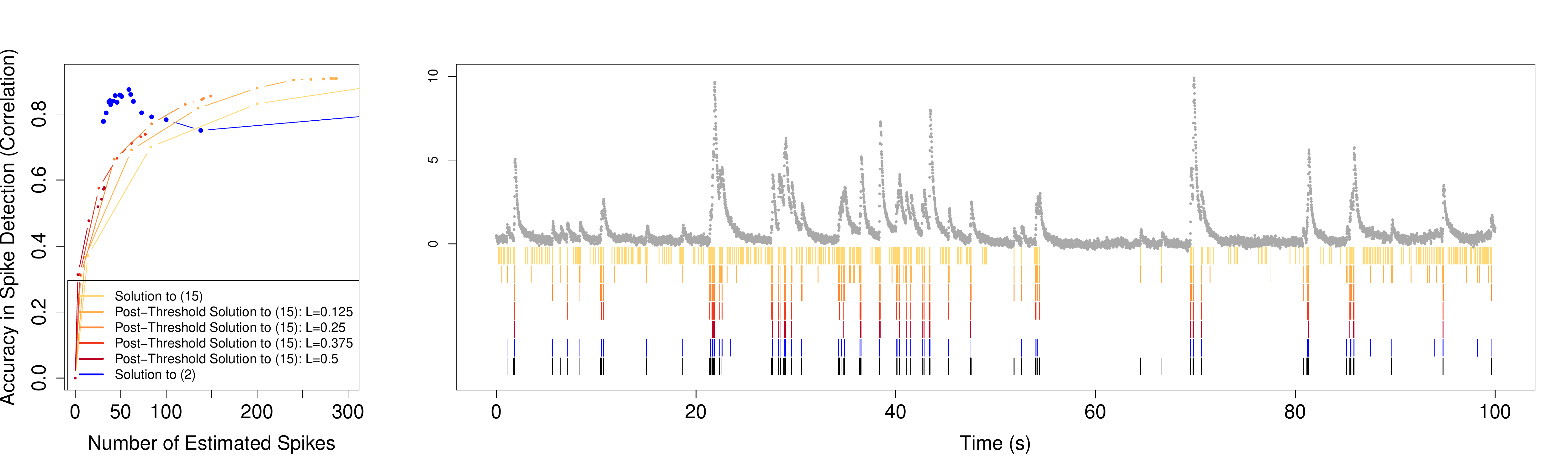}
\caption{Illustrative example for cell 13, GCaMP6f, V1, from \cite{chen2013ultrasensitive} after preprocessing; see \cite{theis2016benchmarking}. Different spike measures are displayed in each row. \textit{Left:} Performances of the post-thresholded solution to \eqref{eq:convex-fried} and the solution to \eqref{eq:nonconvex-pos}. \textit{Right:} The cell's fluorescence trace is displayed in grey. The estimated spikes on the test set from the ``best'' choice of the tuning parameter $\lambda$, as determined by either van Rossum, Victor-Purpura, or a correlation-based measure on the training set, are displayed under the fluorescence trace. The true spike times, as determined using electrophysiological recording, are shown in black. The colors in the left-hand panels correspond to the colors in the right-hand panel.}
\label{fig:example-distances-comparison}
\end{center}
\end{figure}
\end{landscape}
\restoregeometry
\subsubsection{Results for all datasets in the \texttt{spikefinder} challenge} 
\label{sec:results-spikefinder}

In this section, we examine the performance of the solutions to \eqref{eq:nonconvex-pos} and \eqref{eq:convex-fried} on all datasets collected as part of the \texttt{spikefinder} challenge. For the ten datasets included in this challenge, Table~\ref{ta:spikefind} tabulates the calcium indicator, circuit, publishing authors, average, minimum, and maximum fluorescence trace length, the number of cells measured, and the time-scale classification. In total, there are 174 traces, each of which contains fewer than 100,000 timesteps. We analyze these 174 cells as described in Section~\ref{sec:exp-methods-desc}.
 
\begin{table}[ht]
\centering
\caption{Datasets collected from the \texttt{spikefinder} challenge. All datasets were resampled to 100 Hz as part of the challenge.}
\label{ta:spikefind}
\begin{tabular}{lllrrrrl}
  \hline
Indicator & Circuit & Authors & Mean T & Min T & Max T & Num. Cells & Time-scale\\ 
  \hline
OGB-1 & V1 & Theis et al. 2016 & 67598 & 35993 & 71986 & 11 & Medium\\ 
  OGB-1 & V1 & Theis et al. 2016 & 32285 & 9682 & 35508 & 21 & Medium\\ 
  GCamp6s & V1 & Theis et al. 2016 & 43637 & 16802 & 53229 & 13 & Slow \\ 
  OGB-1 & Retina & Theis et al. 2016 & 27763 & 27763 & 27763 & 6 & Medium\\ 
  GCamp6s & V1 & Theis et al. 2016 & 16919 & 16919 & 16919 & 9 & Slow\\ 
  GCaMP5k & V1 & Akerboom et al. 2012 & 19331 & 5998 & 23998 & 9 & Medium\\ 
  GCaMP6f & V1 & Chen et al. 2013 & 23910 & 21715 & 23973 & 37 & Fast\\ 
  GCaMP6s & V1 & Chen et al. 2013 & 23402 & 11986 & 23973 & 21 & Slow \\ 
  jRCAMP1a & V1 & Dana et al. 2016 & 29460 & 6486 & 31959 & 20 & Slow\\ 
  jRGECO1a & V1 & Dana et al. 2016 & 26086 & 5507 & 31994 & 27 & Fast \\ 
   \hline
\end{tabular}
\end{table}

\begin{landscape}
\begin{figure}[h!]
\begin{center}
\includegraphics[scale = .95]{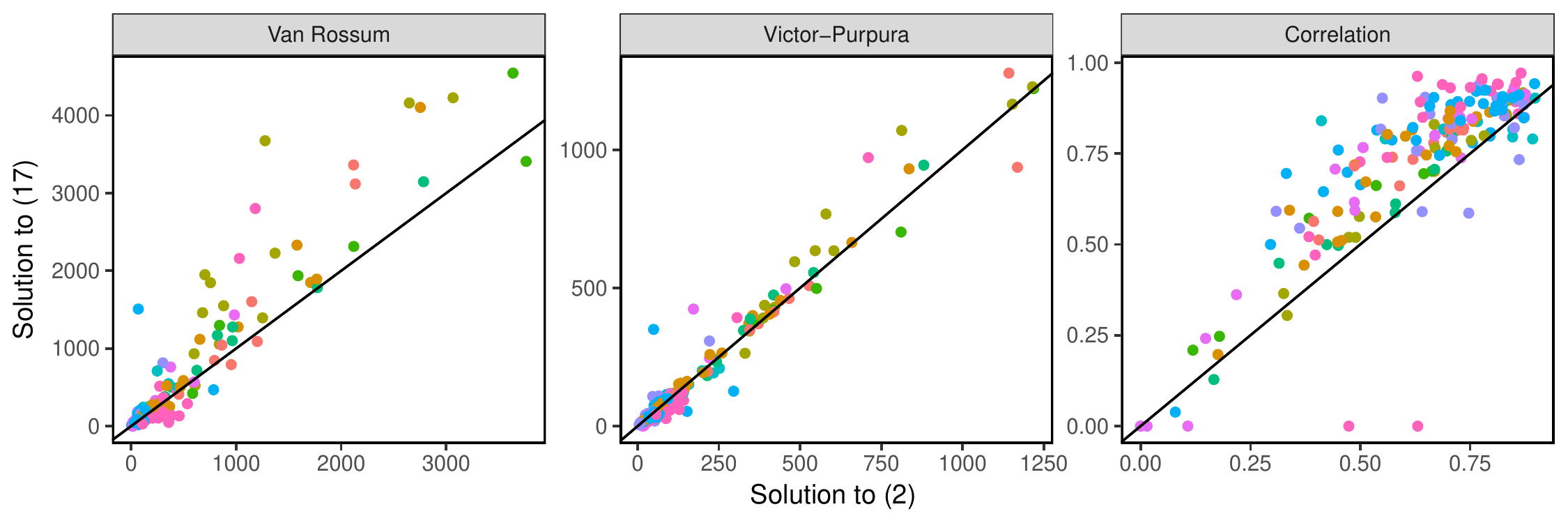}
\caption{Optimal van Rossum, Victor-Purpura, and correlation measures for our proposal, \eqref{eq:nonconvex-pos}, and a competing proposal, \eqref{eq:convex-fried}. Small values of the van Rossum and Victor-Purpura measures suggest accurate estimation of the timing and number of spikes, whereas a large value of the correlation measure suggests accurate estimation of the spike rate, though perhaps an overestimate of the number of spikes. Each dot represents the performance of \eqref{eq:nonconvex-pos} and \eqref{eq:convex-fried} on a single cell, for each of the 174 cells. Cells are colored based on the dataset from which they were obtained (see Table~\ref{ta:spikefind}). }
\label{fig:compare-spikefinder}
\end{center}
\end{figure}
\end{landscape}

Figure~\ref{fig:compare-spikefinder} compares the test set performance, with respect to the van Rossum, Victor-Purpura, and correlation measures, for each of the 174 cells. As measured by the van Rossum and Victor-Purpura distance, the solution to \eqref{eq:nonconvex-pos} outperforms the solution to \eqref{eq:convex-fried}. However, under the correlation measure, the solution to \eqref{eq:convex-fried} achieves higher correlations than the solution to \eqref{eq:nonconvex-pos}. These results are consistent with those on a single cell presented in Section~\ref{sec:single-cell}, in which it was shown that van Rossum and Victor-Purpura are able to quantify spike accuracy and timing, whereas correlation provides a cruder measure of spike rate and encourages over-estimation of the number of spikes.

\subsection{The solution to \eqref{eq:nonconvex-pos} outperforms the solution to \eqref{eq:nonconvex-nopos}} 
\label{sec:no-neg-spikes}

\begin{figure}[h!]
\begin{center}
\includegraphics[scale = 0.35]{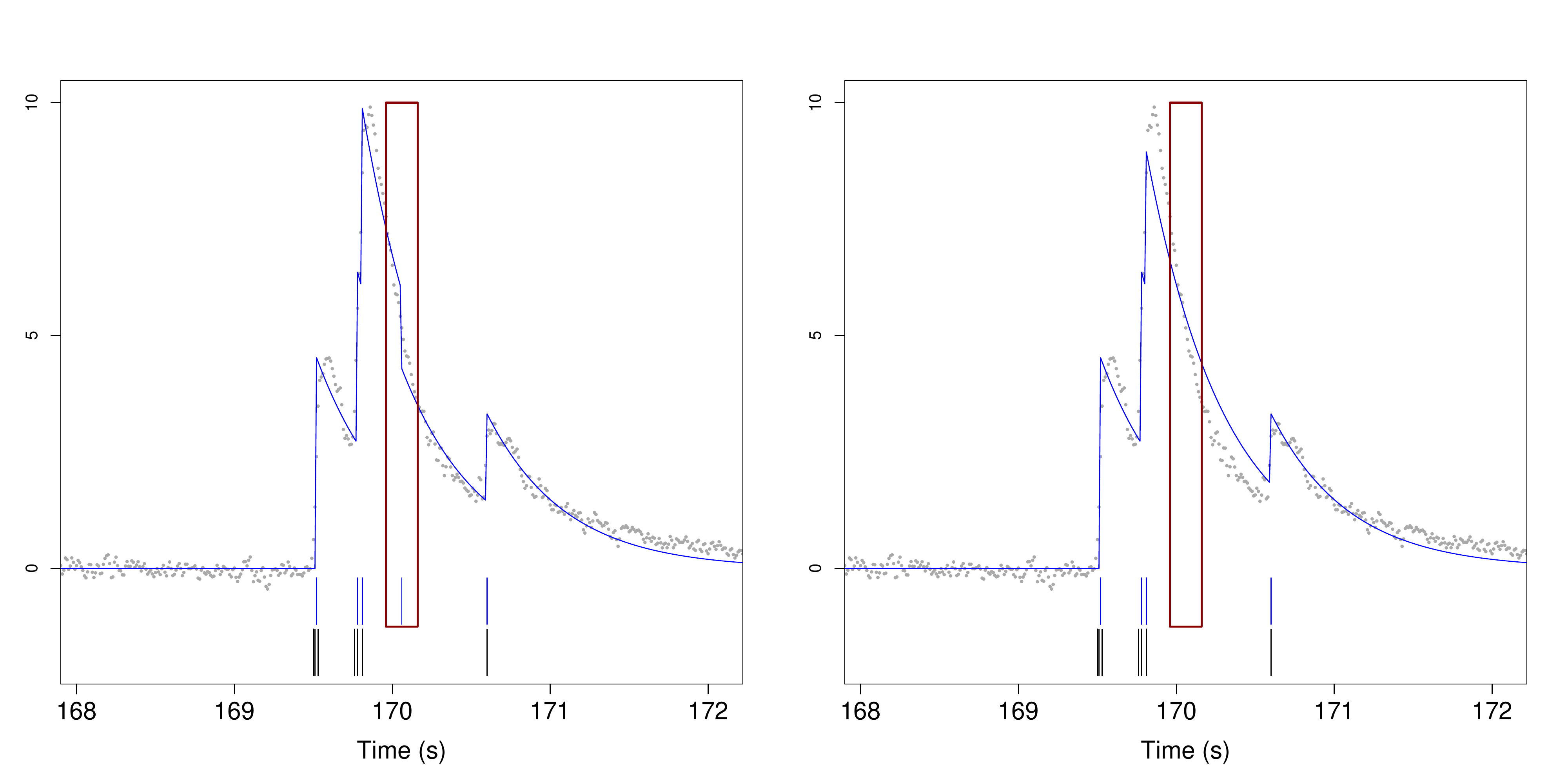}
\caption{Illustrative example to show that solving \eqref{eq:nonconvex-pos} yields better estimates than solving \eqref{eq:nonconvex-nopos}. Fluorescence and spike train data is from cell 13, GCaMP6f, V1, of \cite{chen2013ultrasensitive} after preprocessing; see \cite{theis2016benchmarking}. \textit{Left:} Fluorescence trace and true spikes, as well as the calcium and spikes estimated from \eqref{eq:nonconvex-nopos}. Calcium and estimated spikes are in blue, and true spikes are in black. \textit{Right:} Fluorescence trace and true spikes, as well as  calcium and spikes estimated from \eqref{eq:nonconvex-pos}. Calcium and estimated spikes are in blue, and true spikes are in black. Tuning parameters set to yield the true number of spike times on the full 200s recording. The red box highlights the negative spike in the solution to \eqref{eq:nonconvex-nopos}. }
\label{fig:no-neg-spikes}
\end{center}
\end{figure}

As mentioned earlier, in this paper we have developed not only an algorithm for solving \eqref{eq:nonconvex-nopos} that is much faster than the algorithm proposed in \cite{jewell2017exact}, but also an algorithm for solving \eqref{eq:nonconvex-pos}, which cannot be solved using techniques from \cite{jewell2017exact}. By incorporating the fact that a firing neuron causes an increase, but never a decrease, in the calcium concentration, the estimated spikes from problem \eqref{eq:nonconvex-pos} are closer to the ground truth spikes than the estimated spikes from \eqref{eq:nonconvex-nopos}. Figure~\ref{fig:no-neg-spikes} displays the estimated calcium and spike times for the solutions to \eqref{eq:nonconvex-nopos} and \eqref{eq:nonconvex-pos} for a short illustrative time window, on just one cell, with tuning parameters set to yield the true number of spikes on the full 200s recording. We see that the solution to \eqref{eq:nonconvex-nopos} yields a ``negative'' spike at time 170.06s, that is, $c_{170.06s} - \gamma c_{170.05s} < 0$. By contrast, the solution to \eqref{eq:nonconvex-pos} avoids ``negative'' spikes.

\subsection{Calibrating the estimated spike magnitudes from \eqref{eq:nonconvex-pos} with the number of observed spikes}
\label{sec:calibrate}

The datasets described in Table~\ref{ta:spikefind} are resampled at 100Hz; therefore, a neuron may spike more than once during a single timestep. In these instances, we expect larger increases in the calcium concentration than if a neuron only fired once. We now investigate whether there is a relationship between the magnitude of $\hat{c}_{(\hat{\tau}_{i} + 1)} - \gamma \hat{c}_{\hat{\tau}_{i} }$ and the number of spikes observed at timestep $(\hat{\tau}_{i} + 1)$.

\begin{figure}[h!]
\begin{center}
\includegraphics[scale = 0.6]{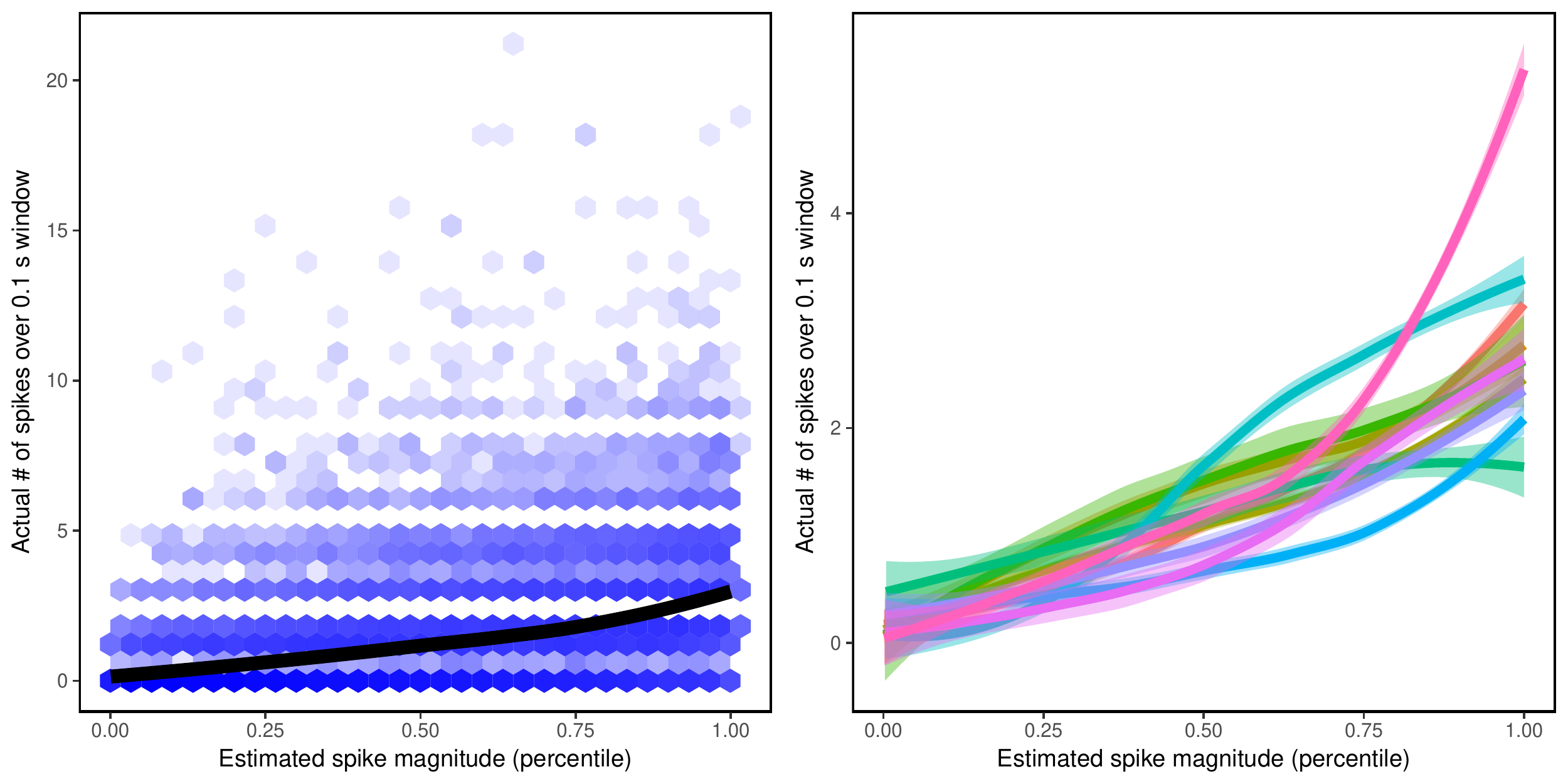}
\caption{Large increases in the estimated calcium $\hat{c}_{t} - \gamma \hat{c}_{t-1}$ are associated with more spikes in the ground truth data. \textit{Left:} Black loess curve fit to all ten datasets. Color intensity denotes the number of estimated spikes within a hexagonal bin. \textit{Right:} Loess curves along with $95\%$ confidence intervals for each dataset. Cells are colored based on the dataset from which they were obtained (see Table~\ref{ta:spikefind} and Figure~\ref{fig:compare-spikefinder}). Details are provided in Section~\ref{sec:calibrate}.}
\label{fig:spike-mag}
\end{center}
\end{figure}

Because the magnitude of $\hat{c}_{(\hat{\tau}_{i} + 1)} - \gamma \hat{c}_{\hat{\tau}_{i} }$ is not directly comparable across data sets, for each cell we transform the magnitudes into percentiles. We then compare the percentile of $\hat{c}_{(\hat{\tau}_{i} + 1)} - \gamma \hat{c}_{\hat{\tau}_{i} }$ to the number of spikes within a 0.1 second window of $(\hat{\tau}_i + 1)$. Figure~\ref{fig:spike-mag} displays the percentiles and the number of spikes across all 174 traces on a test set; tuning parameters were chosen to optimize the van Rossum distance on a training set. The left panel displays a loess curve fit to all ten datasets, and the right panel shows the loess curves along with  $95\%$ confidence intervals for each dataset. As expected, a larger value of  $\hat{c}_{(\hat{\tau}_{i} + 1)} - \gamma \hat{c}_{\hat{\tau}_{i} }$ is associated with more spikes in the ground truth data. 

\section{Discussion}
\label{sec:discussion}

Determining the times at which a neuron fires from a calcium imaging dataset is a challenging and important  problem. In this paper, we  build upon the nonconvex approach for spike deconvolution proposed in \cite{jewell2017exact}. Though \cite{jewell2017exact} proposed a tractable algorithm for solving the nonconvex problem, it is prohibitively slow to run on large populations of neurons for which long recordings are available. The algorithm proposed in this paper solves the optimization problem of \cite{jewell2017exact} for fluorescence traces of 100,000 timesteps in less than a second. Moreover, Algorithm~\ref{alg:fpop-minless} overcomes a limitation of \cite{jewell2017exact} by avoiding ``negative'' spikes; that is, a decrease in the calcium concentration due to a spike. In Section~\ref{sec:real-data}, we show that these algorithms have superior performance, relative to existing approaches, as quantified by the van Rossum and Victor-Purpura measures, on datasets collected as part of the \texttt{spikefinder} challenge (\url{http://spikefinder.codeneuro.org/}). 

Our \texttt{C++} implementation, along with \texttt{R} and \texttt{python} wrappers, is publicly available on \texttt{Github} at \url{https://github.com/jewellsean/FastLZeroSpikeInference}.

\section*{Acknowledgments}
We thank Michael Buice, Peter Ledochowitsch, and Michael Oliver at the Allen Institute for Brain Science and Ilana Witten at Princeton for helpful conversations. Sean Jewell received funding from the Natural Sciences and Engineering Research Council of Canada. Toby Hocking is partially supported by the Natural Sciences and Engineering Council of Canada grant RGPGR 448167-2013, and by Canadian Institutes of Health Research grants EP1-120608 and EP1-120609. This work was partially supported by Engineering and Physical Sciences Research Council Grant EP/N031938/1 to Paul Fearnhead, and NIH Grant DP5OD009145 and NSF CAREER Award DMS-1252624 to Daniela Witten. 

\clearpage
\appendix
\section{Proof of Proposition~\ref{prop:recursion}}
\label{sec:fun-proofs}

The main tool to prove Proposition~\ref{prop:recursion} is a simple recursion for the cost function of segmenting data $y_{1:s}$ with most recent changepoint $\tau$,  for $\tau<s$
\begin{align}
\mathrm{Cost}_s^\tau(\alpha) &=  F(\tau) + \frac{1}{2} \sum_{t=\tau + 1}^{s} \left( y_t - \alpha \gamma^{t - s}\right)^2 +\lambda \nonumber \\
&=\min_{\alpha'} \mathrm{Cost}_\tau^*(\alpha') + \frac{1}{2} \sum_{t=\tau + 1}^{s} \left( y_t - \alpha \gamma^{t - s}\right)^2 +\lambda \nonumber \\
&= \min_{\alpha'} \mathrm{Cost}_\tau^*(\alpha') + \frac{1}{2} \sum_{t=\tau + 1}^{s-1} \left( y_t - \alpha \gamma^{t - s}\right)^2  + \lambda + \frac12 (y_{s} - \alpha)^{2} \nonumber \\
&=  \mathrm{Cost}_{s-1}^\tau(\alpha/\gamma) + \frac12(y_s - \alpha)^2,
\label{eq:basic-functional-recursion}
\end{align}
and, for $\tau = s$, we define 
\begin{align}
\mathrm{Cost}_{s-1}^{s-1}(\alpha/\gamma) \equiv \min_{\alpha'} \mathrm{Cost}_{s-1}^*(\alpha')+\lambda.
\label{eq:cost-tau-equal-s}
\end{align}
Therefore, given $\mathrm{Cost}_{s-1}^{\tau}(\alpha)$, we can form $\mathrm{Cost}_{s}^{\tau}(\alpha)$. 


We now turn our attention to the optimal cost functions $\mathrm{Cost}_{s}^{*}(\alpha)$. Consider splitting the minimization over changepoints, $0\leq \tau < s$, in  $\mathrm{Cost}_{s}^{*}(\alpha)$ as the minimum over changepoints at timesteps $\tau < s-2$ and at $s-1$,
\begin{align*}
\mathrm{Cost}_{s}^*(\alpha) &= \min_{0\leq \tau < s}\mathrm{Cost}_{s}^\tau(\alpha) = \min\left\{ \min_{\tau < s-2}\mathrm{Cost}_{s}^\tau(\alpha), \mathrm{Cost}_{s}^{s-1}(\alpha)\right\}, 
\end{align*}
which combined with the simple recursion in \eqref{eq:basic-functional-recursion} gives
\begin{align*}
\mathrm{Cost}_{s}^*(\alpha) &= \min\left\{ \min_{\tau < s-2}\mathrm{Cost}_{s-1}^\tau(\alpha/\gamma) + \frac12(y_s - \alpha)^2, \mathrm{Cost}_{s-1}^{s-1}(\alpha/\gamma) + \frac12(y_s - \alpha)^2 \right\}  \\
&= \min\left\{ \mathrm{Cost}_{s-1}^*(\alpha/\gamma), \min_{\alpha'} \mathrm{Cost}_{s-1}^*(\alpha'/\gamma) + \lambda \right\} + \frac12(y_s - \alpha)^2 \\
&= \min\left\{ \mathrm{Cost}_{s-1}^*(\alpha/\gamma), \min_{\alpha'} \mathrm{Cost}_{s-1}^*(\alpha') + \lambda \right\} + \frac12(y_s - \alpha)^2.
\end{align*}

The second-to-last equality results from the fact that $\mathrm{Cost}_{s-1}^{*}(\alpha/\gamma)=\min_{\tau<s-2} \mathrm{Cost}_{s-1}^\tau(\alpha/\gamma)$ (see \eqref{eq:cost-star}), and from \eqref{eq:cost-tau-equal-s}. This completes the proof of Proposition~\ref{prop:recursion}.

\clearpage
\section{A fast functional pruning algorithm for problem \eqref{eq:nonconvex-pos}}
\label{sec:fast-algo-min-less}

Algorithm~\ref{alg:fpop-minless} efficiently solves \eqref{eq:nonconvex-pos}. We notice that Algorithm~\ref{alg:fpop-minless} is almost identical to Algorithm~\ref{alg:fpop}, with the exception to line 2. However, there is also a subtle difference as $\mathrm{Cost}_{s}^{\tau}(\alpha)$ is defined as in \eqref{eq:cost-tau-constrain}, whereas in Algorithm~\ref{alg:fpop}, $\mathrm{Cost}_{s}^{\tau}(\alpha)$ is defined as in \eqref{eq:cost-tau}.

\begin{algorithm}[h!]
    \SetKwInOut{Initialize}{Initialize}
    \SetKwInOut{Output}{Output}
	\Initialize{Compute $\mathrm{Cost}_{1}^{*}(\alpha) := \frac12 (y_{1} - \alpha)^{2}$}
	\ForEach{timestep $s = 2, \ldots, T$}{
		Calculate and store $\mathrm{Cost}_{s}^{*}(\alpha) := \min\{ \mathrm{Cost}_{s-1}^{*}(\alpha / \gamma), \min_{\alpha \geq \alpha'}\mathrm{Cost}_{s-1}^{*}(\alpha' / \gamma) + \lambda\} + \frac12 (y_{s} - \alpha)^{2}$}
		
		// Use optimal cost functions to determine changepoints $\tau_{1}, \ldots, \tau_{k}$\\
		Initialize list of changepoints $cp := (T)$ \\
		Set the current changepoint $\tau^{cur} := T$ \\
		Initialize list of estimated calcium concentrations $c := ()$ \\
			\While{$\tau^{cur} > 0$}{
$\tau^{prev} := \tau^{cur}$ \\
Determine the most recent changepoint $\tau^{cur} := \{ \tau : \min_{\alpha}	\mathrm{Cost}_{\tau^{prev}}^{\tau}(\alpha) = \min_{\alpha} \mathrm{Cost}_{\tau^{prev}}^{*}(\alpha)		\}$ \\
Determine the calcium concentration at $\tau^{prev}$, $\alpha^{*} := \argmin{\alpha}{\mathrm{Cost}_{\tau^{prev}}^{\tau^{cur}}(\alpha)} $ \\
Update list of calcium concentrations, $ c := (\alpha^{*}, c)$ \\
\ForEach{timestep $s = (\tau^{prev} - 1), \ldots, (\tau^{cur} + 1)$}{
Calculate calcium concentration, $\alpha^{*} / \gamma$, and then append to list, $c := (\alpha^{*} / \gamma, c)$ \\
Scale $\alpha^{*} := \alpha^{*} / \gamma$ \\
}

Update list of changepoints $cp := (\tau^{cur}, cp)$ \\
		} 
    \Output{Set of changepoints $cp$, number of changepoints $k := \mathrm{card}(cp)$, and estimated calcium concentrations $c$.
    } 
\caption{A functional pruning algorithm for solving \eqref{eq:nonconvex-pos} }\label{alg:fpop-minless}
\end{algorithm}

\clearpage
\section{Implementation considerations} 
\label{sec:implementation} 

The recursive updates in \eqref{eq:min-less-cost} can give rise to numerical overflow. To see why this occurs, recall from \eqref{eq:piecewise} that $\mathrm{Cost}_s^*(\alpha)$ is a piecewise polynomial. For $\alpha \in \mathcal{R}_s^\tau$, computing $\mathrm{Cost}_s^*(\alpha)$ through the recursion in \eqref{eq:min-less-cost} requires computing $\mathrm{Cost}_t^*(\alpha)$ for $t=\tau,\tau+1,\ldots,s$; each of these computations requires scaling the quadratic coefficient by $1/\gamma^2$, and equivalently, scaling $\alpha$ by a factor of $\gamma$. This means that in order to obtain $\mathrm{Cost}_s^*(\alpha)$, the quadratic coefficients in $\mathrm{Cost}_\tau^*(\alpha)$ are scaled by a factor of $(1/\gamma^2)^{\tau-s}$. For $\tau-s$ large, this is a large number, leading to numerical overflow.

The issue of numerical overflow is particularly pronounced when $\mathrm{Cost}_s^*(\alpha)$ in \eqref{eq:piecewise} is a function of $\mathcal{R}_s^\tau$ for $\tau \ll s$; this tends to occur when no true spikes have occurred in the recent past before the $s$th timestep, or when the value of $\lambda$ is very large. In light of these observations, in implementing \eqref{eq:nonconvex-pos}, we exclude any regions $\mathcal{R}_s^\tau$ from \eqref{eq:piecewise} that are bounded above by a very small positive constant $\rho=10^{-40}$. This can be seen as approximately constraining $c_t \geq \rho$, or else limiting the amount of time that can elapse between consecutive changepoints.



\section{Example of recursion \eqref{eq:min-less-cost} for solving \eqref{eq:nonconvex-pos}} 
\label{sec:min-less-example}

\begin{example} \label{ex:recursion-minless}
We continue Example~\ref{ex:recursion}, but this time we use \eqref{eq:min-less-cost} to solve \eqref{eq:nonconvex-pos} rather than using \eqref{eq:min-cost} to solve \eqref{eq:nonconvex-nopos}. Once again, consider the simple dataset $y = [1.00, 0.98, 0.96, \ldots]$ with $\lambda = \frac12$ and $\gamma = 0.98$. We start with $\mathrm{Cost}_{1}^{*}(\alpha)$, which is just the quadratic centered around $y_{1}$, 
\begin{align*}
\mathrm{Cost}_{1}^{*}(\alpha) = \mathrm{Cost}_{1}^{0}(\alpha) = \frac12 (y_{1} - \alpha)^{2} = \frac12 (1.00 - \alpha)^{2}.
\end{align*}

To calculate $\mathrm{Cost}_{2}^{*}(\alpha)$ based on \eqref{eq:min-less-cost}, we first calculate 
\begin{align*}
\min_{\alpha' \leq \alpha} \mathrm{Cost}_{1}^*(\alpha'/\gamma) = 
\begin{cases}
  \frac12 (1 - \alpha / \gamma)^{2}, & 0 \leq \alpha < \gamma \\
0, & \alpha \geq \gamma
\end{cases}.
\end{align*}
We then use \eqref{eq:min-less-cost} to calculate
\begin{align*}
\mathrm{Cost}_{2}^{*}(\alpha) &= 
\min\left\{ \mathrm{Cost}_{1}^*(\alpha/\gamma), \min_{\alpha' \leq \alpha} \mathrm{Cost}_{1}^*(\alpha'/\gamma) + \lambda \right\} + \frac12(y_2 - \alpha)^2  \\
&= 
\begin{cases}
\frac12( 1 - \alpha / \gamma)^{2} + \frac12(0.98 - \alpha)^2, &  \alpha \in \mathcal{R}_{2}^{0}\equiv[0, 2\gamma) \\
\frac12 + \frac12(0.98 - \alpha)^2, &  \alpha \in \mathcal{R}_{2}^{1}\equiv[2\gamma, \infty).
\end{cases}
\end{align*}

Again, to obtain $\mathrm{Cost}_{3}^{*}(\alpha)$, we calculate
\begin{align*}
\min_{\alpha' \leq \alpha} \mathrm{Cost}_{2}^*(\alpha'/\gamma) = 
\begin{cases}
\frac12( 1 - \alpha / \gamma^{2})^{2} + \frac12(0.98 - \alpha/\gamma)^2, &  \alpha \leq \gamma^{2} \\
0, & \alpha \geq \gamma^{2}
\end{cases}.
\end{align*}
Then using the recursion \eqref{eq:min-less-cost} we obtain the optimal cost function
\begin{align*}
\mathrm{Cost}_{3}^{*}(\alpha) &= 
\min\left\{ \mathrm{Cost}_{2}^*(\alpha/\gamma), \min_{\alpha' \leq \alpha} \mathrm{Cost}_{2}^*(\alpha'/\gamma) + \lambda \right\} + \frac12(y_3 - \alpha)^2 \\
&=  \frac12(0.96 - \alpha)^2 + 
\begin{cases}
\frac12( 1 - \alpha / \gamma^{2})^{2} + \frac12(0.98 - \alpha/\gamma)^2,  & \alpha \in \mathcal{R}_{3}^{0} \equiv \left(0, \gamma^{2}\left( 1 + \frac{1}{\sqrt{1 + \gamma^{2}}}\right)\right)\\
 \lambda, & \alpha \in   \mathcal{R}_{3}^{2} \equiv \left[\gamma^{2}\left( 1 + \frac{1}{\sqrt{1 + \gamma^{2}}}\right), \infty\right)
\end{cases}.
\end{align*}

Figure~\ref{fig:evolve-min-less} illustrates these updates. This example illustrates that the optimal cost $\mathrm{Cost}_{3}^{*}(\alpha)$ corresponding to \eqref{eq:nonconvex-pos} differs from \eqref{eq:nonconvex-nopos}. In particular, the region corresponding to $\mathcal{R}_{3}^{0}$ from \eqref{eq:nonconvex-pos} is $\left(0, \gamma^{2}\left( 1 + \frac{1}{\sqrt{1 + \gamma^{2}}}\right)\right)$, whereas region corresponding to $\mathcal{R}_{3}^{0}$ from \eqref{eq:nonconvex-nopos} is $\gamma^{2}\left\{\left[0,  1 - \frac{1}{\sqrt{1 + \gamma ^{ 2}}} \right) \cup \left[ 1 + \frac{1}{\sqrt{1 + \gamma ^{ 2}}}, \infty\right)\right\}$; compare Figures~\ref{fig:evolve-min-less} and \ref{fig:evolve}. However, since the optimal value of $\mathrm{Cost}_{3}^{*}(\alpha)$ occurs with most recent changepoint zero, the solutions to problems \eqref{eq:nonconvex-pos} and \eqref{eq:nonconvex-nopos} are identical.

 \begin{figure}[htbp]
\begin{center}
\begin{tabular}{lc}
 \begin{minipage}{0.07\textwidth}$s = 1:$\end{minipage}  &  \begin{minipage}{0.949\textwidth}\includegraphics[scale = 0.38]{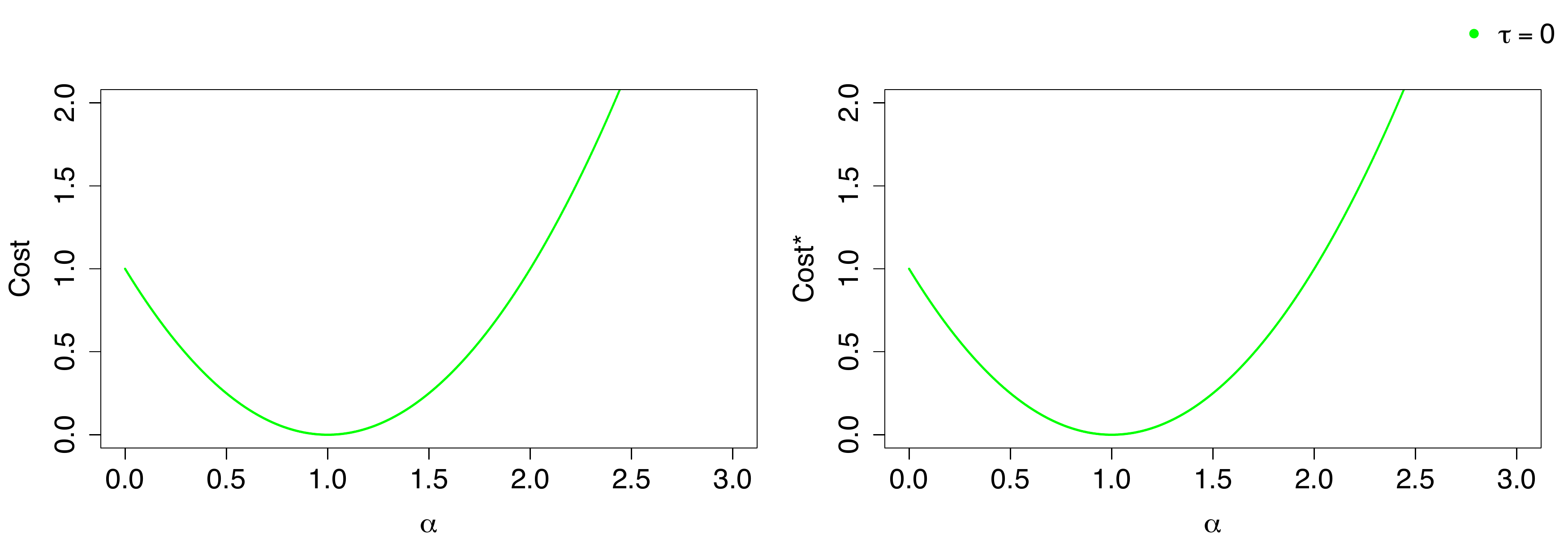} \end{minipage} \\
 \begin{minipage}{0.07\textwidth}$s = 2:$\end{minipage}  &  \begin{minipage}{0.949\textwidth}\includegraphics[scale = 0.38]{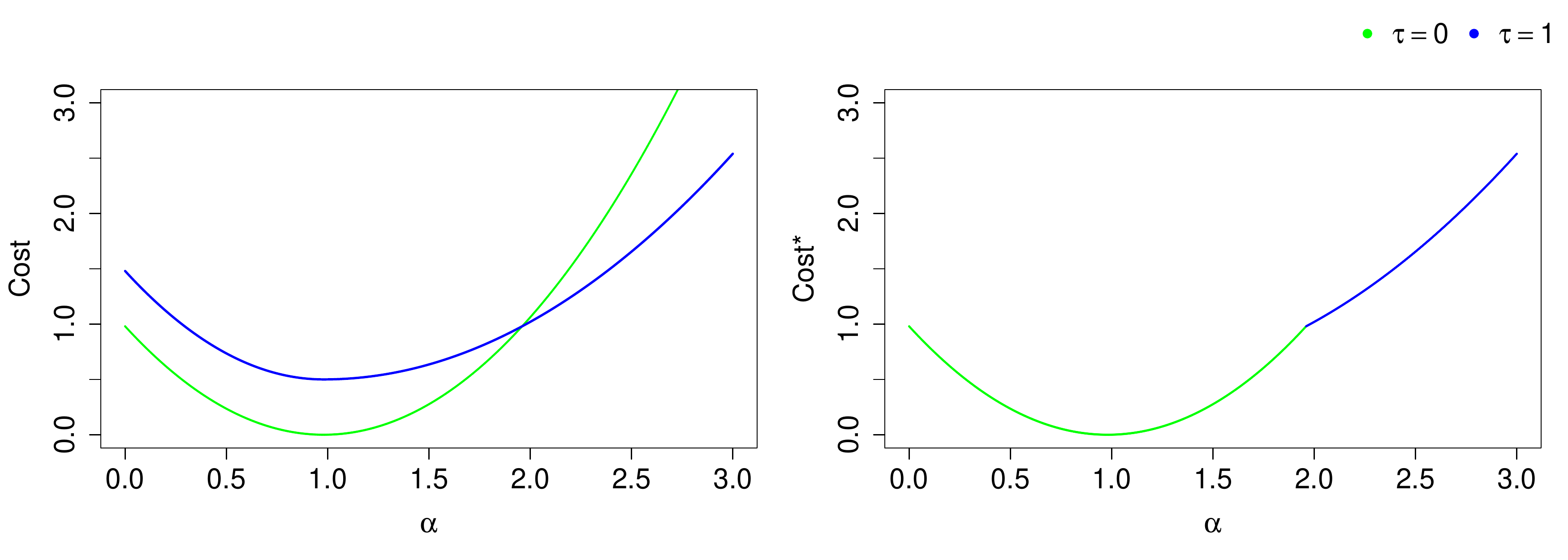} \end{minipage} \\
 \begin{minipage}{0.07\textwidth}$s = 3:$\end{minipage}  &  \begin{minipage}{0.949\textwidth}\includegraphics[scale = 0.38]{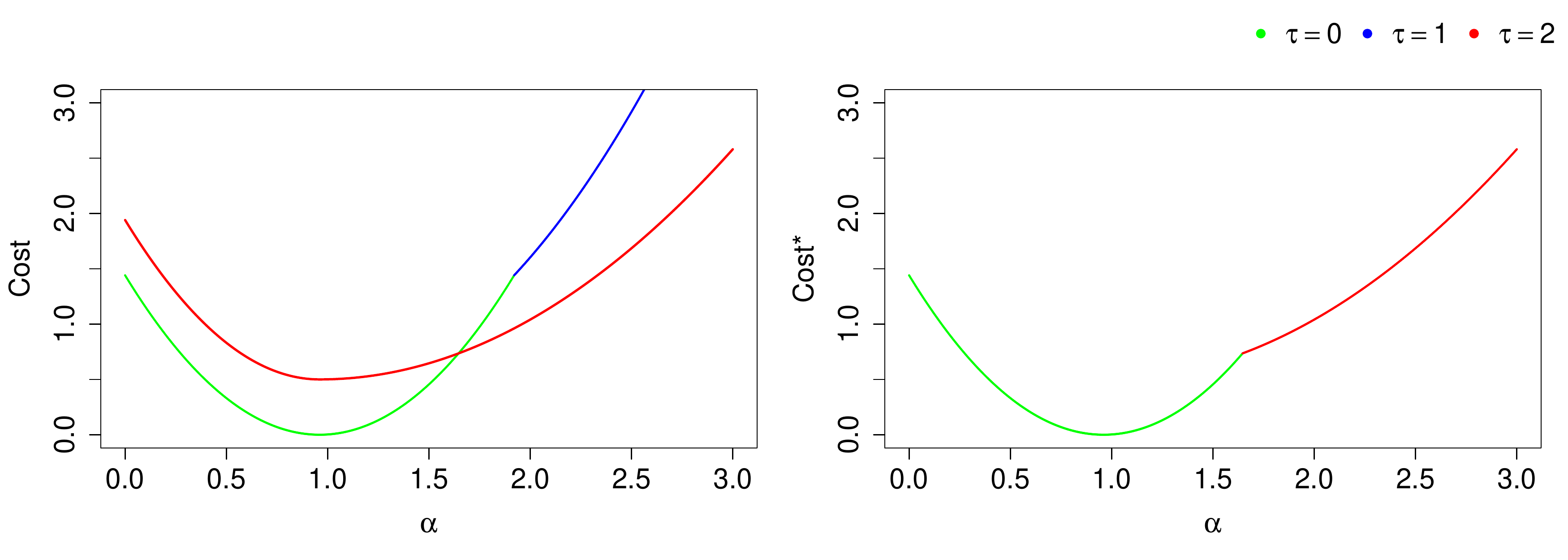} \end{minipage} \\
\end{tabular}
\caption{Evolution of $\mathrm{Cost}_{s}^{\tau}$ and $\mathrm{Cost}_{s}^{*}(\alpha)$ for Example~\ref{ex:recursion-minless}. The left-hand panels display the functions $\mathrm{Cost}_{s-1}^*(\alpha/\gamma) + \frac12(y_s - \alpha)^2$ and $\min_{\alpha'\leq \alpha} \mathrm{Cost}_{s-1}^*(\alpha') + \lambda  + \frac12(y_s - \alpha)^2$, and the right-hand panels show the function $\mathrm{Cost}_{s}^{*}(\alpha)$, which is the minimum of those two functions. Rows index the timesteps, $s = 1, 2, 3$. The functions are colored based on the timestep of the most recent changepoint, that is, the value of $\tau$ corresponding to $\mathcal{R}_s^{\tau}$. \emph{Top:} When $s=1$, $\mathrm{Cost}_1^*(\alpha)=\frac{1}{2}(1-\alpha)^2$; this corresponds to the region $\mathcal{R}_1^0 = [0, \infty)$. \emph{Center:} When $s=2$, $\mathrm{Cost}_2^*(\alpha)$ is the minimum of two quantities: $\mathrm{Cost}_{1}^*(\alpha/\gamma) + \frac12(y_2 - \alpha)^2$, which corresponds to the most recent changepoint being at timestep zero, and $\min_{\alpha'\leq \alpha} \mathrm{Cost}_{1}^*(\alpha') + \lambda + \frac12(y_2 - \alpha)^2$, which corresponds to the most recent changepoint being at timestep one. These two functions are shown on the left-hand side, and $\mathrm{Cost}_2^*(\alpha$) is shown on the right-hand side. \emph{Bottom:} When $s=3$, $\mathrm{Cost}_3^*(\alpha)$ is calculated similarly; see Example~\ref{ex:recursion-minless} for additional details.}
\label{fig:evolve-min-less}
\end{center}
\end{figure}

 \end{example}

\bibliographystyle{imsart-nameyear}
\bibliography{ms}

\end{document}